\documentclass[longbibliography, twocolumn, secnumarabic, amssymb, nobibnotes, superscriptaddress, aps, prx, amsmath]{revtex4-2}

\usepackage{graphicx}
\usepackage{subfigure}
\usepackage{appendix}
\usepackage[ruled, linesnumbered]{algorithm2e}
\usepackage{booktabs}
\usepackage{color}
\usepackage{bm}
\usepackage{soul}
\usepackage{makecell}
\usepackage{threeparttable}
\usepackage{hyperref}
\hypersetup{colorlinks=true, citecolor=blue, urlcolor=blue, linkcolor=blue}
\soulregister{\cite}{7}
\soulregister{\ref}{7}

\begin{document}

\title{Performing SU($d$) operations and rudimentary algorithms in a superconducting transmon qudit for $d=3$ and $d=4$}

\author{Pei Liu}
\email{liu-p20@mails.tsinghua.edu.cn}
\affiliation{
State Key Laboratory of Low Dimensional Quantum Physics,
Department of Physics,
Tsinghua University,
Beijing 100084,
China}

\author{Ruixia Wang}
\affiliation{
Beijing Academy of Quantum Information Sciences,
Beijing 100193,
China
}

\author{Jing-Ning Zhang}
\email{zhangjn@baqis.ac.cn}
\affiliation{
Beijing Academy of Quantum Information Sciences,
Beijing 100193,
China
}

\author{Yingshan Zhang}
\affiliation{
Beijing Academy of Quantum Information Sciences,
Beijing 100193,
China
}

\author{Xiaoxia Cai}
\affiliation{
Beijing Academy of Quantum Information Sciences,
Beijing 100193,
China
}

\author{Huikai Xu}
\affiliation{
Beijing Academy of Quantum Information Sciences,
Beijing 100193,
China
}

\author{Zhiyuan Li}
\affiliation{
Beijing Academy of Quantum Information Sciences,
Beijing 100193,
China
}

\author{Jiaxiu Han}
\affiliation{
Beijing Academy of Quantum Information Sciences,
Beijing 100193,
China
}

\author{Xuegang Li}
\affiliation{
Beijing Academy of Quantum Information Sciences,
Beijing 100193,
China
}

\author{Guangming Xue}
\affiliation{
Beijing Academy of Quantum Information Sciences,
Beijing 100193,
China
}

\author{Weiyang Liu}
\email{liuwy@baqis.ac.cn}
\affiliation{
Beijing Academy of Quantum Information Sciences,
Beijing 100193,
China
}

\author{Li You}
\affiliation{
State Key Laboratory of Low Dimensional Quantum Physics,
Department of Physics,
Tsinghua University,
Beijing 100084,
China}
\affiliation{
Beijing Academy of Quantum Information Sciences,
Beijing 100193,
China
}
\affiliation{
Frontier Science Center for Quantum Information,
Beijing 100184, 
China
}

\author{Yirong Jin}
\affiliation{
Beijing Academy of Quantum Information Sciences,
Beijing 100193,
China
}

\author{Haifeng Yu}
\affiliation{
Beijing Academy of Quantum Information Sciences,
Beijing 100193,
China
}

\date{\today}

\begin{abstract}

Quantum computation architecture based on $d$-level systems,
or qudits,
has attracted considerable attention recently due to their enlarged Hilbert space.
Extensive theoretical and experimental studies have addressed aspects of algorithms and benchmarking techniques for qudit-based quantum computation and quantum information processing.
Here, 
we report a physical realization of qudit with upto 4 embedded levels in a superconducting transmon,
demonstrating high-fidelity initialization,
manipulation,
and simultaneous multi-level readout.
In addition to constructing SU($d$) operations and benchmarking protocols for quantum state tomography,
quantum process tomography,
and randomized benchmarking etc,
we experimentally carry out these operations for $d=3$ and $d=4$.
Moreover,
we perform prototypical quantum algorithms and observe outcomes consistent with expectations.
Our work will hopefully stimulate further research interest in developing \textcolor{black}{manipulation protocols and efficient applications} for quantum processors with qudits.

\end{abstract}

\maketitle

\section{Introduction}

Quantum computational advantage is largely enabled by the exponentially growing Hilbert space for storing and processing of information.
In the most commonly discussed architecture,
the basic unit is a two-level system,
forming a qubit,
with the computation space growing \textcolor{black}{as $2^N$} for $N$ qubits.
\textcolor{black}{This exponential scaling can be further extended to $d^N$ by introducing qudits,
i.e. quantum $d$-level systems,
as basic computational units \cite{gottesman1999fault, wang2020qudits}.
Such an expanded Hilbert space can be realized without increased hardware complexity in popular quantum-computation platforms \cite{Ringbauer2022, erhard2018twisted}.
Besides a larger Hilbert space and saved hardware resource,
other potential advantages of qudits have also attracted considerable research interest.}
For example,
the accuracy and efficiency of simple quantum circuits and algorithms can be enhanced by qudit-based architecture \cite{chi2022programmable}.
Quantum simulation can enjoy the flexibility of qudits,
with which the many-body Hamiltonian can be encoded directly for simulating higher spin systems \cite{gonzalez2022hardware},
such as bosonic spin-1 models with $d=3$.
In quantum-nonlocality-based information processing, 
the qudit also plays an important role by helping to close out the detection loophole often tormenting Bell test experiments \cite{vertesi2010closing, lo2016experimental}.
In quantum key distribution,
the qudit can lead to increased security and higher key rate \cite{cerf2002security},
and with the qudit as a quantum repeater,
the improved communication scheme is possible \cite{bergmann2019hybrid}.

Implementations of qudits have been studied on various physical platforms.
For trapped ions with multi-levels,
an experimental realization is reported recently \cite{shen2017quantum}.
A universal operation set for implementing qudit-based computation,
including state preparation,
single-qudit gates,
two-qudit gates,
and measurement schemes is provided \cite{low2020practical, Ringbauer2022}.
Qudits based on multi-level atom arrays are employed to explore dipole-dipole interactions \cite{munro2018population}.
For photonic systems,
a variety of inherent properties of a photon,
including its orbital angular momentum \cite{babazadeh2017high},
frequency-bin \cite{imany2019high, kues2017chip},
time-bin \cite{islam2017provably, humphreys2013linear},
and path \cite{chi2022programmable},
have been used to construct qudits.
Moreover,
qudit-based quantum computation is studied in continuous spin systems \cite{adcock2016quantum},
NV centers in diamond \cite{fu2022experimental},
and nuclear-magnetic-resonance (NMR) systems \cite{choi2017dynamical, dogra2014determining}.
In recent years,
qubit-based superconducting quantum computation (SQC) research has witnessed significant progresses in  
quantum machine learning \cite{huang2022quantum},
quantum chemistry \cite{malley2016scalable, huggins2022unbiasing},
quantum simulation \cite{satzinger2021realizing},
quantum error correction \cite{krinner2022realizing},
and quantum computational advantage with more than 50 qubits \cite{arute2019quantum} etc,
with transmon as a favored physical realization due to its insensitivity to charge noise \cite{koch2007charge}.
\textcolor{black}{Due to its tunable multi-level structure,
superconducting transmon is naturally made for implementing qudit,
and investigating the manipulations of higher-excited states has become a significant priority to realize qudit-based architecture for quantum information processing.
In qubit-based SQC where quantum information is encoded in computational space unit spanned by the lowest two levels,
higher-excited states also play a non-negligible role in implementation of quantum operations,
such as two-qubit gate \cite{strauch2003quantum, rigetti2010fully, chu2023implenting} and shelving readout \cite{englert2010mesoscopic, elder2020high} protocols.}

\textcolor{black}{Recently,
employing superconducting transmon qudits as basic units for quantum information processing has attracted increased attention,
with quantum information encoded into an expanded Hilbert space augmented by higher-excited states.
Operational protocols for universal gates and algorithms have been explored theoretically for transmon qudit \cite{kiktenko2015multilevel}.
However,
most earlier studies focus on qutrit manipulation for $d=3$ \cite{bianchetti2010control, cervera2022experimental, brown2022trade, blok2021quantum, xu2016coherent, luo2022experimental, goss2022high} or simulation for $d=4$ \cite{zheng2022optimal},
limited by the nature of the transmon.
The coherence time of higher-excited states decrease as the number of excitations increases,
approximately proportional to $1/m$ with $m$ labeling the number of excitations \cite{peterer2015coherence}.
The charge parity effect,
which often manifests itself as a beat note in the Ramsey interference,
causes an increased frequency dichotomy for higher levels \textcolor{black}{\cite{peterer2015coherence, tennant2022low}},
and hence undermines precise manipulations through frequency addressing or phase correction.
The implementation of qudit-based architecture thus calls for stricter requirements on the quality of transmon device.
Specifically,
it requires the transmon to have longer coherence time and weaker charge-parity effect.
When both of these two requirements are fulfilled,
as it is for the device used in our experiment,
advantages of qudits emerge.
In addition,
the ability to simultaneously discriminate multiple states also contributes to high-fidelity implementation.}

\textcolor{black}{This work implements high-fidelity qudit manipulations for $d=3$ and $d=4$ in a specifically designed and fabricated superconducting transmon exhibiting long coherence time and weak charge-parity effect.
We accomplish simultaneous $4$-state readout with fidelity above $91.1\%$ for each state.
To benchmark the performance,
we prepare and measure a $4$-level state with the fidelity of $99.64\%$,
and experimentally estimate the error per gate as $(7.6\pm0.1)\times10^{-4}$ ($(1.5\pm0.1)\times10^{-3}$) for $\pi/2$ pulses between levels $\{|1\rangle, |2\rangle\}$ ($\{|2\rangle, |3\rangle\}$).
We also implement several rudimentary algorithms to show the efficacy and efficiency of the single-qudit processor.
Our experiments demonstrate the feasibility of encoding and processing more than one bit of quantum information in a single superconducting transmon,
and we hope it will stimulate more interest in theoretical and experimental studies on the qudit-based quantum information processing architecture.}

This paper is arranged as follows.
First,
the arbitrary SU($d$) operation construction with its physical realization and common benchmarking protocols,
such as quantum state tomography \textcolor{black}{\cite{bonk2004quantum, bianchetti2010control}},
quantum process tomography,
and randomized benchmarking \cite{Ringbauer2022} are discussed in section~\ref{gate}.
Then discrete Fourier transform algorithm, 
Grover's algorithm,
and variational quantum eigensolver algorithm in quantum chemistry are implemented in our qudit system as applications in section~\ref{application}.
\textcolor{black}{We end with conclusion and outlook in section~\ref{outlook}.}

\section{Gate Design and Benchmarking\label{gate}}

In this section,
we describe the \textcolor{black}{gate decomposition theory} and the relevant protocols for benchmarking.
The subsections will respectively cover the construction protocols for SU($d$) operations with their physical realizations in a transmon qudit in~\ref{construction},
and protocols for quantum state tomography,
quantum process tomography,
and randomized benchmarking in~\ref{benchmarking}.

\subsection{Construction of SU($d$) Operations\label{construction}}

\begin{figure*}
    \includegraphics[width=7in]{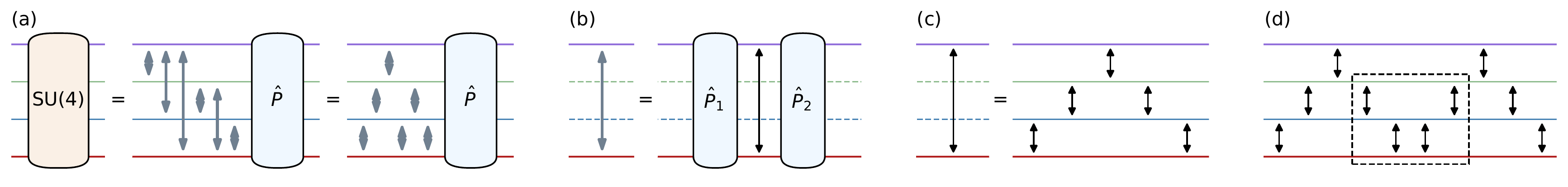}
    \caption{\label{fig:circuit}
        A circuit diagram illustration for arbitrary unitary operations on a four-level system.
        (a) The Gaussian elimination method decomposes a SU(4) operation into a sequence of SU(2) operations plus a generalized phase gate,
        as in normal or bubbling Gaussian elimination.
        Here gray arrows represent SU(2) operations in a subspace spanned by any two levels.
        (b) Such SU(2) operations can be constructed by two generalized phase gates with a two-level rotation operation $\hat R_{m, n}(\theta, \phi)$ sandwiched in between.
        Black arrows represent transitions between two levels with a microwave drive.
        (c) The arbitrary rotation $\hat R_{m, n}(\theta, \phi)$ can be realized by direct multi-photon coupling, 
        or a sequence of transitions between adjacent levels without crossing to nonadjacent ones.
        (d) If a transition $\hat R_{m, n}(\theta', \phi')$ follows another $\hat R_{m, n}(\theta, \phi)$ in the same two-level subspace,
        \textcolor{black}{some auxiliary adjacent transitions according to Eq.~(\ref{r=rrr})},
        as shown in the black dashed box,
        \textcolor{black}{can be neglected at all time},
        which constitutes a straightforward strategy to simplify circuit.
    }
\end{figure*}

Microwave-driven transitions can drive a qudit,
connecting levels in a quantum system as in trapped ion system or superconducting one etc,
with designated single- or multi-photon transitions to perform qudit manipulations.
Between any two of $d$ levels,
the coupled transition provides a universal form
\begin{equation}
    \hat R_{m, n}(\theta, \phi)=\exp\left[-\frac{i\theta}2\left(\cos\phi\hat\sigma_x^{m,n}+\sin\phi\hat\sigma_y^{m,n}\right)\right] \label{rmn}
\end{equation}
of rotation,
with $\hat\sigma_x^{m,n}=|m\rangle\langle n|+|n\rangle\langle m|$ and $\hat\sigma_y^{m,n}=-i|m\rangle\langle n|+i|n\rangle\langle m|$.
Here,
$|m\rangle$ and $|n\rangle$ with $m,n=0, 1, \cdots, d-1$ and $m\neq n$ are arbitrary basis states of the $d$ level qudit.
With $\hat R_{m, n}(\theta, \phi)$,
transitions that cannot be directly implemented are constructed from a combination of several transitions in sequence,
for example,
\begin{equation}
    \begin{aligned}
        \hat R_{m, m+2}(\theta, \phi)=&\hat R_{m, m+1}(-\pi, \phi_1)\\
        &\hat R_{m+1, m+2}(\theta, \phi_2)\hat R_{m, m+1}(\pi, \phi_1), \label{r=rrr}
    \end{aligned} 
\end{equation}
with $\phi_1+\phi_2-\pi/2=\phi$,
i.e.,
from first swapping the amplitudes between $|m\rangle$ and $|m+1\rangle$,
then carrying out the rotation operation between $|m+1\rangle$ and $|m+2\rangle$,
and finally swapping the amplitudes for $|m+1\rangle$ and $|m\rangle$ back.
To simplify the above composite operation,
we typically choose $\phi_2=\phi$ and $\phi_1=\pi/2$ as a symmetric and standard protocol.
Whenever $\hat R_{m, n}(\theta,\phi)$ precedes another rotation $\hat R_{m, n}(\theta',\phi')$ on the same two levels,
it is convenient to eliminate some constituting pulses in the sandwiched structure of Eq.~(\ref{r=rrr}) to simplify the total sequence.

Besides microwave-driven transitions,
the generalized phase gate,
defined as $\hat P(\vec\Phi)=\sum_k\exp\left(i\phi_k\right)|k\rangle\langle k|$,
is another important repertoire for qudit manipulation.
It can be realized by $3(d-1)$ resonant pulses according to Ref. \cite{Ringbauer2022}.
Often it gets too complicated,
particularly when it appears in the middle of a sequence as an independent unitary operation.
Inspired by the idea of virtual Z gate from qubit control \cite{mckay2017efficient} and noting that
\begin{equation}
    \hat R_{m, n}(\theta, \phi)\hat P(\vec\Phi)=\hat P(\vec\Phi)\hat R_{m, n}(\theta, \phi+\phi_m-\phi_n), \label{pr=rp}
\end{equation}
we find that one can swap the generalized phase gate from arbitrary positions of a circuit to the very beginning,
and then ignore it,
as with the virtual Z gate \cite{mckay2017efficient}.
Such a virtual operation requires zero time,
hence can be executed perfectly.
With such a generalized virtual phase gate,
operation sequences can be further simplified.
An arbitrary SU($2$) operation can then be constructed effectively as in the following,
\begin{equation}
    \begin{aligned}
        \hat U_2^{m, n}(\theta, \phi, \lambda, \delta)&=\hat P(\vec\Phi^{(1)})\hat R_{m, n}\left(\theta, \frac\pi2\right)\hat P(\vec\Phi^{(2)}),\label{rf}\\
        \phi_{k}^{(1)}&=\begin{cases}
            -\frac\lambda2, &k=m,\\
            \frac\lambda2, &k=n,\\
            0, &{\rm{otherwise}},
        \end{cases},\\
        \phi_{k}^{(2)}&=\begin{cases}
            \delta-\frac\phi2, &k=m,\\
            \delta+\frac\phi2, &k=n,\\
            0, &{\rm{otherwise}},
        \end{cases}.
    \end{aligned}
\end{equation}

The actual decomposition of an arbitrary unitary operation on a $d$-level qudit into a sequence of unitary operations on two-level subsystems follows the idea of Gaussian elimination \cite{barenco1995elementary, nielsen2010, oleary2006parallelism}.
SU($d$) operation can normally be expressed as a unitary matrix of order $d$ and the Gaussian elimination algorithm takes this matrix as input,
and outputs a sequence of SU(2) operations between various pair of the $d$ levels,
or simply expressed as in the following
\begin{equation}
	\hat U_d^{0, 1, \cdots, d-1}=\hat P(\vec\Phi)\hat U_2^{m_k, n_k}\cdots\hat U_2^{m_1, n_1}\hat U_2^{m_0, n_0}, \label{gauss}
\end{equation}
where a total of $(k+1)$ SU(2) operations are required in a given order and $m_l, n_l\in\{0, 1, \cdots, d-1\}, l=0, 1, \cdots, k$.
There are two strategies of Gaussian elimination.
The normal one and the bubbling one are described in Appendix~\ref{app:gauss} Algorithm~\ref{alg:gauss} and Algorithm~\ref{alg:gauss2},
respectively.
Figure~\ref{fig:circuit} illustrates the decomposition process for $d=4$.
With \textcolor{black}{the} abovementioned protocol,
an arbitrary SU($d$) operation can be constructed based on two-level ones

\begin{figure}
    \includegraphics[width=3.4in]{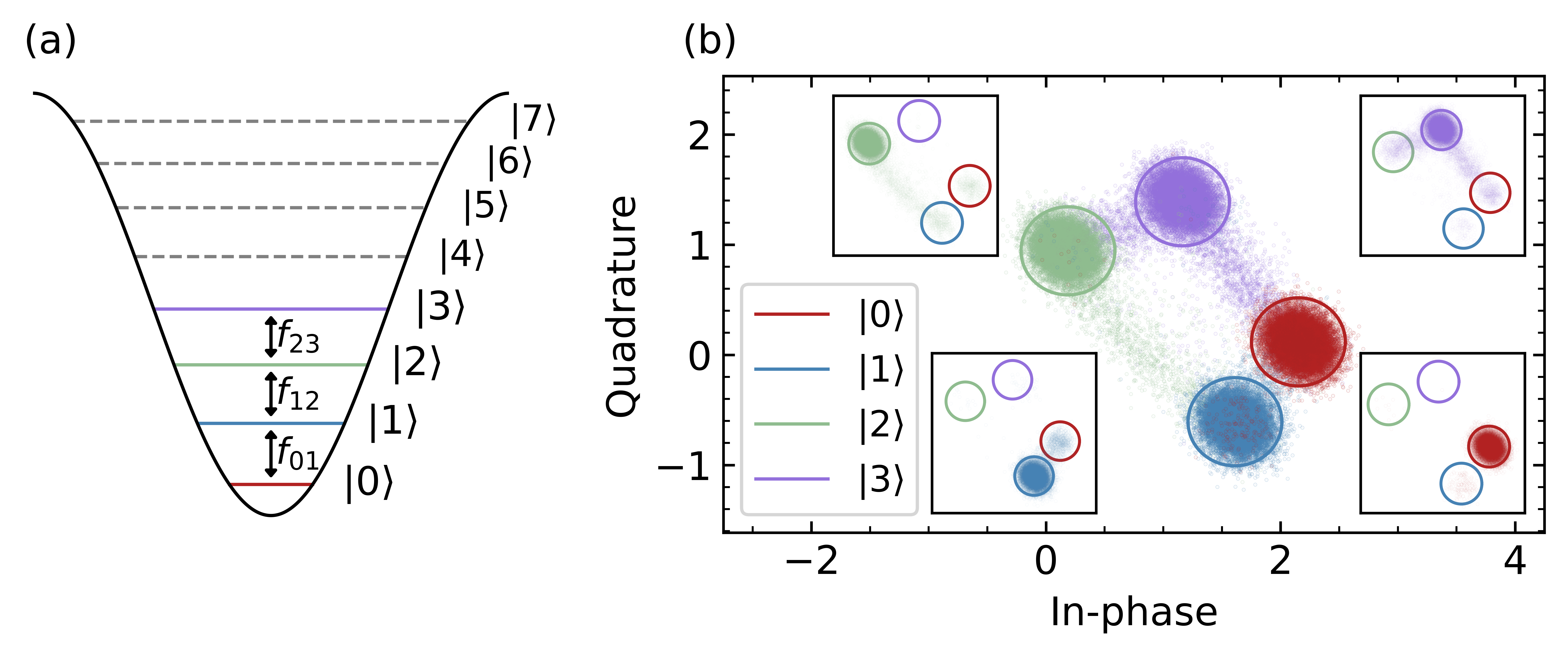}
    \caption{\label{fig:device}
        Qudit level diagram and readout calibration.
        (a) Quantum state is encoded as a superposition of several bound levels in a cosine potential well,
        labeled by $|0\rangle$,
        $|1\rangle$,
        $|2\rangle$,
        and $|3\rangle$.
        Higher levels further up of this device are plotted in gray dashed lines.
        (b) The quadrature representation of readout benchmarking,
         with fidelities $P_0=99.1\%$,
         $P_1=94.5\%$,
         $P_2=94.5\%$,
         and $P_3=91.1\%$,
         qudit prepared in $|k\rangle$ with a possible global phase and then readout $|k\rangle$ after the measurement process with a probability $P_k$.
    }
\end{figure}

\textcolor{black}{As for experiments,}
our qudit system is constructed with a superconducting transmon \cite{koch2007charge},
whose Hamiltonian can be expressed as $\hat H=4E_C(\hat n-n_g)^2-E_J\cos\hat\phi$ with $E_c$ the charging energy and $E_J$ the Josephson energy.
For suitable parameters,
several bound states exist in the cosine potential well,
as shown in Fig.~\ref{fig:device}(a).
The anharmonicity of the cosine potential for the qudit ensures transition frequencies between any two energy levels are different.
Several single- or two-photon transitions are shown in Appendix~\ref{fig:drive} with specific parameters for our device in Table~\ref{tab:error}.
It is easy to find that for adjacent transitions,
or transitions between neighboring levels of a transmon,
shorter evolution time is needed than for two-photon ones due to stronger single-photon couplings.
\textcolor{black}{To suppress charge parity effect,
$E_J/E_c$ is designed to take a large ratio ($E_J/E_C\approx88$) compared to the usual values. 
}

The transmon qudit is coupled to a readout resonator, 
\textcolor{black}{whose frequency} responses differently when the qudit is prepared in different states,
facilitating simultaneous readout of the qudit directly.
As shown in Fig.~\ref{fig:device}(b),
the first four states labeled by $|0\rangle$,
$|1\rangle$,
$|2\rangle$,
and $|3\rangle$ can be distinguished with high fidelity.
\textcolor{black}{The details on the readout can be found in Appendix~\ref{app:params}}.

Taking into considerations of both energy level structure and readout fidelity,
we can implement a qudit system of $d\le4$ with three transitions on neighboring levels over short operation time,
although two-photon transitions as well as the fifth state $|4\rangle$ ($d=5$) remain observable which would likely increase our ability to manipulate for $d>4$ in the future.
The beat frequency of Ramsey interference between $|2\rangle$ and $|3\rangle$ is less than $0.1\ \rm{MHz}$,
thus only impairs the manipulations slightly which constitutes the key reason for our success in the high-fidelity SU(4) manipulations.
At the same time,
if we choose the bubbling Gaussian elimination instead,
which is the most suitable for the qudit under discussion,
no more than $d(d-1)/2$ operations are needed,
resulting in at most $d(d-1)$ $\pi/2$ pulses and a complexity of $\mathcal O(d^2)$ that is close to the theoretical limit.
\textcolor{black}{
Mathematically Gaussian elimination is not restricted to a specific elimination order.
In other words,
both the normal one and the bubbling one can fulfill the same goal.
But the order of elimination,
or the elimination strategies,
exhibit different complexities in a specific quantum system depending on the level structure.
We choose normal Gaussian elimination which is widely known and general in this work,
leading to a complexity of $\mathcal O(d^3)$,
when an arbitrary SU(2) operation has to be expanded in terms of adjacent SU(2) transitions.
This strategy avoids optimizations for a specific system than for the bubbling one,
especially in the present system,
consistent with our motivation for a universal demonstration.}

\subsection{Quantum State Tomography, Quantum Process Tomography, and Randomized Benchmarking\label{benchmarking}}

\begin{figure}
    \includegraphics[width=3.4in]{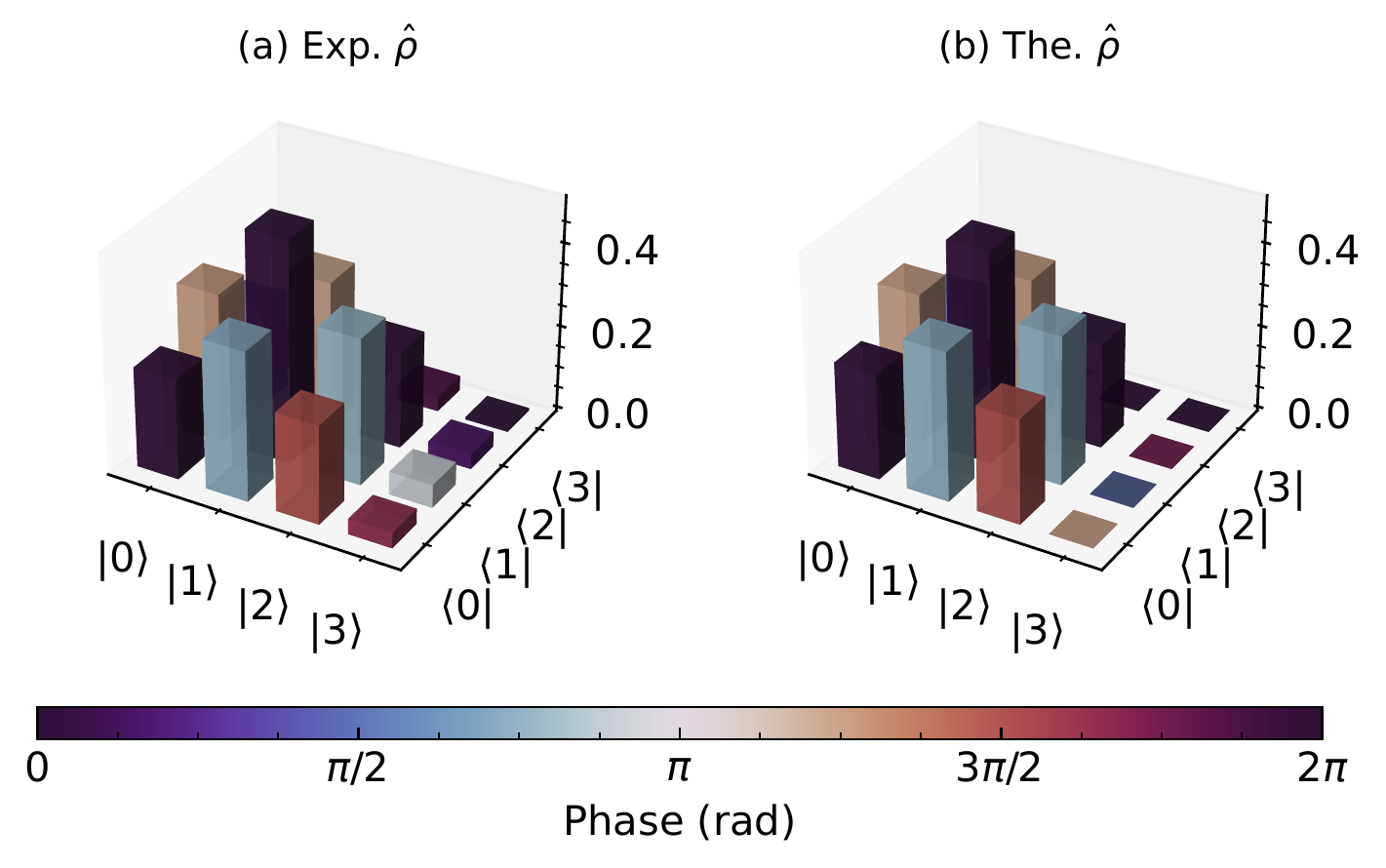}
    \caption{\label{fig:qst}
		Quantum state tomography with MLE for the state $|\psi\rangle=(1-i)|0\rangle/\sqrt8+|1\rangle/\sqrt2-(1+i)|2\rangle/\sqrt8$ in the four-level qudit,
        where (a) shows experimental results and (b) shows the expected theoretical ones.
        Height of each bar represents amplitude while color represents phase.
        The corresponding state fidelity is $99.64\%$.
    }
\end{figure}

\begin{figure}
    \includegraphics[width=3.4in]{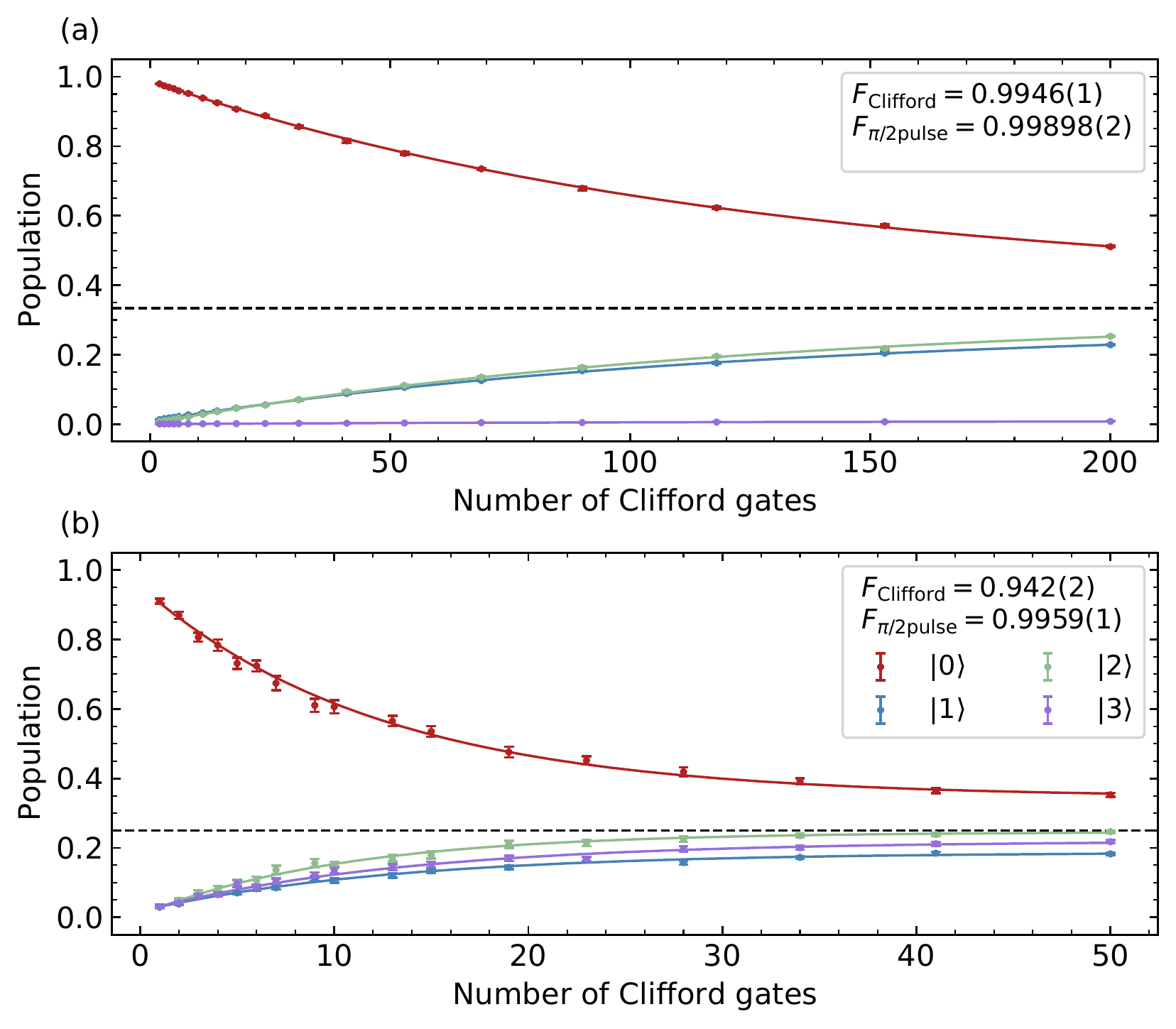}
    \caption{\label{fig:rb34}
        Randomized benchmarking for (a) three-level and (b) four-level qudits.
		We initialize the qudit in $|0\rangle$,
		then apply a certain extra number of Clifford gates from (a) $3$-dimensional Clifford group $\mathcal C_3$ (involving $|0\rangle$, $|1\rangle$, and $|2\rangle$) and (b) $4$-dimensional Clifford group $\mathcal C_4$ (involving $|0\rangle$, $|1\rangle$, $|2\rangle$, and $|3\rangle$), 
	    and finally add a Clifford gate to ensure that the total sequence is equivalent to identity. 
        Populations of the four states $|0\rangle$,
        $|1\rangle$,
        $|2\rangle$,
        and $|3\rangle$ are measured simultaneously in the end,
        which are shown as crosses with error bars and corresponding colors.
        Solid lines with corresponding colors show exponential fits according to the theory of RB. 
        $\pi/2$ pulses with an average number of $A_3=5.25$ are used in $\mathcal C_3$ and average number of $A_4=14.292$ in $\mathcal C_4$,
        which give the average calculated fidelities of $\pi/2$ pulses as $F_{\pi/2\rm{pulse}}=1-(1-F_{\rm{Clifford}})/A_k, k=3,4$.
        Black dashed lines indicate the populations of a mixed state after a sufficiently long evolving time.
    }
\end{figure}

Quantum state tomography (QST) is a standard method for determining the density matrix of a state.
Here we follow the multi-level QST protocol developed in NMR system \cite{bonk2004quantum} \textcolor{black}{and superconducting transmon qutrit \cite{bianchetti2010control}}.
With the readout process presented in subsection~\ref{construction},
only the diagonal elements of a density matrix operator $\hat\rho$ can be accessed.
To measure an arbitrary density matrix element of a qudit,
several operations for swapping off-diagonal elements to linear combinations of the diagonal ones are needed before measurement.
For $d=4$,
we apply a total of 12 operations $\hat M_l, l=0, 1, \cdots, 11$ (listed in Appendix~\ref{app:qst}) before measurement,
and the measured probabilities are $P_{l, k}=\langle k|\hat M_l^\dagger\hat\rho\hat M_l|k\rangle$ with $l=0,1, \cdots, 11$ and $k=0, 1, 2, 3$,
and $\sum_k P_{l, k}=1$.
An overdetermined group of equations can then be derived,
whose solution gives the unknown $\hat\rho$.
According to the properties of density matrix,
maximum likelihood estimation (MLE) \cite{myung2003tutorial, knee2018maximum} with the simple estimation of $\hat\rho$ from Eq.~(\ref{rho}) taken as the initial guess,
reduces the impact of other undesirable errors on the output density matrix.
Figure~\ref{fig:qst} displays QST measurement results for a superposition state $|\psi\rangle=(1-i)|0\rangle/\sqrt8+|1\rangle/\sqrt2-(1+i)|2\rangle/\sqrt8$,
finding $\hat\rho=|\psi\rangle\langle\psi|$ with a high $99.64\%$ fidelity.
For dimensions with $d>4$,
we can construct an analogous measurement operator.
For $d=3$,
we can simply truncate the above $d=4$ protocol.

Quantum process tomography (QPT) is based on QST \textcolor{black}{\cite{nielsen2010, Ringbauer2022}}.
It provides a convenient measure to characterize a quantum process.
Similar to QPT for qubit,
we initialize our qudit in a set of given states,
then apply the process we want to determine,
and at the end of the process carry out QST to measure the final state.
For a quantum process represented by
\begin{equation}
    \hat\rho_f=\sum_{k,l}\hat\lambda_k\hat\rho_i\hat\lambda_l^\dagger\chi_{k, l},
\end{equation}
with initial (final) state $\hat\rho_i$($\hat\rho_f$),
$\lambda_l$ ($l=0, 1, \cdots, d^2-1$) is the identity matrix or $d^2-1$ generators of SU($d$) group,
corresponding to the operator $\hat\lambda_l$ (details for its construction are available in Appendix~\ref{app:qpt} Algorithm~\ref{alg:sun}).
In particular, 
when $d=3$, the matrices $\lambda_1, \lambda_2, \cdots, \lambda_8$ are known as the Gell-Mann matrices.
Under this basis,
the superoperator $\hat\chi$ (corresponding to matrix representation $\{\chi_{k, l}\}$),
has desirable properties to guarantee MLE in solving the overdetermined group of equations for QPT.
As an example,
benchmarking of discrete Fourier transformation is illustrated in section~\ref{application}.

\begin{table}
    \centering
    \caption{\label{tab:error}
        Error of Gates
    }
    \begin{tabular}{cccc}
        \toprule
            SU($d$)& levels& Clifford& $\pi/2$ pulse\\
        \midrule
            SU(3)& $\{0, 1, 2\}$& $(5.4\pm0.1)\times10^{-3}$& $(1.02\pm0.02)\times10^{-3}$\\
            SU(4)& $\{0, 1, 2, 3\}$& $(5.8\pm0.2)\times10^{-2}$& $(4.1\pm0.1)\times10^{-3}$\\
            SU(2)& $\{0, 1\}$& $(4.6\pm0.3)\times10^{-4}$& $(2.1\pm0.1)\times10^{-4}$\\
            SU(2)& $\{1, 2\}$& $(1.7\pm0.1)\times10^{-3}$& $(7.6\pm0.1)\times10^{-4}$\\
            SU(2)& $\{2, 3\}$& $(2.3\pm0.1)\times10^{-3}$& $(1.5\pm0.1)\times10^{-3}$\\
        \bottomrule
    \end{tabular}
\end{table}

Randomized benchmarking (RB) on three-level and four-level systems can be used to estimate average gate errors, 
just like RB sequence on a qubit \cite{Ringbauer2022}.
We note that Clifford gate groups for qudit are different from that for a single qubit.
More specifically,
there are $216(768)$ group elements for $d=3(4)$ qudit (see Appendix~\ref{app:clifford} for more details).
We translate the matrix representations of Clifford elements into executable sequences of $\pi/2$ pulses.
The average number of $\pi/2$ pulses required to apply Clifford gate in $3$-dimensional Clifford group $\mathcal C_3$ and $4$-dimensional Clifford group $\mathcal C_4$ are $1134/216=5.25$ and $10976/768\approx14.292$,
respectively.
Experimental results are shown in Fig.~\ref{fig:rb34} and Appendix Fig.~\ref{fig:rb2},
with a summary of errors shown in Table~\ref{tab:error}.
We find the average errors calculated from SU(3) and and SU(4) Clifford groups are larger than those from SU(2).

Several explanations are now in order to help understand the situation.
First,
higher levels in transmon typically exhibit shorter energy relaxation time,
implicating more incoherent errors.
In other words, 
the upper bound of fidelity decreases as the energy level becomes higher \cite{abad2021universal},
consistent with decreasing RB fidelity we observe as the subspace expands to include higher levels.
Charge parity effect becomes worse at higher levels as well and the associated frequency dichotomy causes phase uncertainty.
Second,
leakage is regarded as incoherent error in two-level system whereas as coherent error sometimes in multi-level cases.
In the latter case, 
RB is insensitive to such error,
leading to the situation that $\pi/2$ pulse error calculated from RB in $\mathcal C_3$ or $\mathcal C_4$ is larger than from $\mathcal C_2$,
thus more refined leakage control for $d=3$ and $4$ are needed.
Both energy relaxation and frequency bandwidth with finite-time drive pulses contribute to leakage error beyond two-level subspace.
From Fig.~\ref{fig:rb2},
we can find obvious leakage error.
As a result the level population does not converge to the expected value after evolving for a sufficiently long time. 
Transition frequency crowding makes leakage error worse,
because adjacent transition frequency between $|1\rangle$ and $|2\rangle$ is naturally close to the three-photon transition frequency between $|0\rangle$ and $|3\rangle$.
If amplitude of the driving pulse between $|1\rangle$ and $|2\rangle$ is large enough,
frequency bandwidth for a finite-time driving pulse need to be carefully modified to avoid three-photon transitions.
Finally,
due to the averaging effect of RB sequence,
specific types of errors such as non-Markovian error in operations \cite{blume-kohout2017demonstration, greenbaum2015introduction} could not be observed and only limited manipulation errors can be detected,
leading to decreased realized manipulation fidelity.
Analyzing sources of the above mentioned errors is of great importance for improving manipulation accuracy,
and the limit this can be achieved depends on the development of required theoretical tools in the future.

\section{Applications\label{application}}

In this section,
we present three quantum algorithms that are performed to demonstrate capabilities of our transmon qudit.
Benchmarking methods QST and QPT are employed to verify the relevant processes for the respective algorithms.
The experimental results are found to be in nice agreements with theories,
and they illustrate high efficiency and accuracy of our four-level transmon manipulations.

\subsection{Discrete Fourier Transformation and Cyclic Permutation Parity Check}

\begin{figure}
    \includegraphics[width=3.4in]{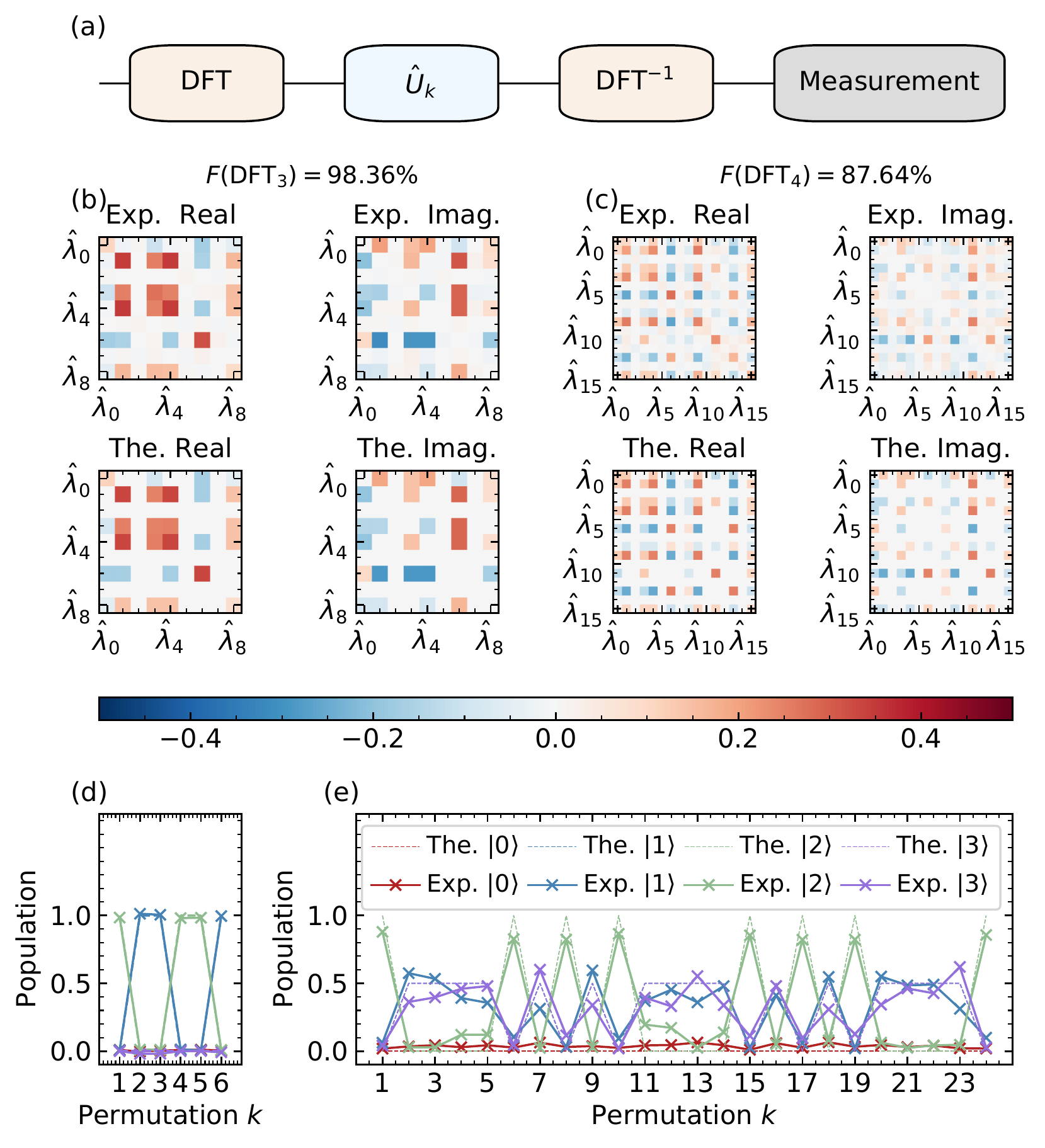}
    \caption{\label{fig:dft}
        Discrete Fourier transformation and parity check of cyclic permutation. 
        (a) The qudit is initialized into a superposition state and the unitary permutation operation is followed by the inverse DFT and readout.
        Quantum process tomography of (b) $\rm{DFT}_3$ and (c) $\rm{DFT}_4$ show fidelities of $98.36\%$ and $87.64\%$ respectively.
        Results for each permutation $k$ (d) for $d=3$ and (e) for $d=4$ with initial state $|2\rangle$ prepared,
        where dashed lines denote theoretical expectations while crosses show total experiment outputs.
        Different colors show the projection measurement results for different states at the same time.
        The consistency of measured population and theoretical distribution affirms the reliable construction of arbitrary unitary operation in transmon qudit,
        though success of the parity checking algorithm depends on the choice of initial state preparation.
        Experimental (theoretical) results in (b)-(e) are labeled by Exp. (The.),
        while the real (imaginary) part are labeled by Real (Imag.).
    }
\end{figure}

The first application we perform is parity check of cyclic permutations using discrete Fourier transformation (DFT).
It has been studied in NMR before \cite{gedik2015computational},
with the circuit diagram as shown in Fig.~\ref{fig:dft}(a).
To check for a permutation of length $d$,
the qudit is initialized into a coherent superposition state $\rm{DFT}_{d}|m\rangle$, $m=0,1,2,\cdots,d-1$.
Then the permutation operation $\hat U_k, k=1, 2, \cdots, d!$ is applied.
Before measurement,
an inverse DFT,
labeled by $\rm{DFT}^{-1}$,
is applied to transform the state into the final one.
Different final state would indicate different parity of cyclic permutation.

Here the process of $\rm{DFT}$ is implemented via the normal Gaussian elimination decomposition,
which translates matrix representation ${\rm{DFT}}_d$ into a sequence of SU(2) operation pulses,
with matrix element ${\rm{DFT}}_d(j, k)$ in row $j$ and column $k$ taking the form
\begin{equation}
    {\rm{DFT}}_d(j, k) = \frac1{\sqrt d}e^{2ijk\pi/d},  j,k=0, 1, \cdots, d-1.
\end{equation}
The operations of ${\rm{DFT}}_3$ and ${\rm{DFT}}_4$ are benchmarked by QPT.
Figures~\ref{fig:dft}(b) and \ref{fig:dft}(c) show the corresponding results,
which give their respective fidelities of $\mathcal F(\rm{DFT}_3)=98.36\%$ and $\mathcal F(\rm{DFT}_4)=87.64\%$.

$\hat U_k$, 
the permutation operator of length $d$,
is given by
\begin{equation}
    \hat U_k=\begin{bmatrix}
        0&1&\cdots&d-1\\
        p_{k, 0}&p_{k, 1}&\cdots&p_{k, d-1}
    \end{bmatrix},
\end{equation}
with $\hat U_k|j\rangle=|p_{k, j}\rangle$,
where $p_{k, j}\in\{0, 1, \cdots, d-1\}, \forall j\in\{0, 1, \cdots, d-1\}$,
and $\forall j_1\neq j_2, p_{k, j_1}\neq p_{k, j_2}$ for $k$ in an ascending lexicographical order of $p_{k, 1}, p_{k, 2}, \cdots, p_{k, d}$.
The simplest construction of $\hat U_k$ uses no more than $d(d-1)/2$ $\pi$ pulses according to properties of permutations,
which constitute an example of pulse sequence optimization for specific operations according to their properties.
More details are provided in Appendix~\ref{app:dft}\ref{app:uk}.

It is worth noting that choice of initial state affects result of this algorithm with $m=0$ being the trivial case and neglected.
When $m$ and $d$ are co-prime numbers,
i.e. $\gcd(m, d)=1$,
the parity of cyclic permutation can be directly obtained from the measurement result of the populations for $|m\rangle$ and $|d-m\rangle$.
The former indicates even parity and the latter affirms odd parity.
If other initial states are chosen,
the results would be a bit complicated.
Suppose $C_{d, \rm{even(odd)}}$ is an arbitrary cyclic permutation of even (odd) parity with length $d$,
and the permutation set $\mathcal G(m, d)$ is a subgroup of a permutation group $\mathcal C(d)$ with length $d$,
which only depends on $m$ and $d$ (more construction details are given in Appendix~\ref{app:dft}\ref{app:parity}).
If a permutation $\hat U_k$ is recognized by the population of state $|m\rangle$ during readout,
it satisfies
\begin{equation}
    \hat U_k\in\bigcup_{g\in\mathcal G(m, d)}g C_{d, \rm{even}},
\end{equation}
while the readout population of state $|d-m\rangle$ corresponds to $C_{d, \rm{odd}}$ with $d=2m$ a special case for which both even and odd permutations give the same readout.

An earlier experiment \cite{gedik2015computational} reported the case of $d=4, m=1$.
Here,
other choices such as $d=3, m=2$ and $d=4, m=2$ are studied, 
with results respectively displayed in Figs.~\ref{fig:dft}(d) and \ref{fig:dft}(e).
The parity check algorithm for the former case is confirmed,
whereas it is not completely established for the latter,
as the result recognizes the cyclic permutation from the whole permutation set but fails in the parity check.
However,
fidelities from QPT of DFT operations and high consistency between measured populations and theoretical distribution of the circuit,
show that arbitrary unitary operations within transmon qudit are realizable. 

\subsection{Grover's Algorithm}

\begin{figure}
    \includegraphics[width=3.4in]{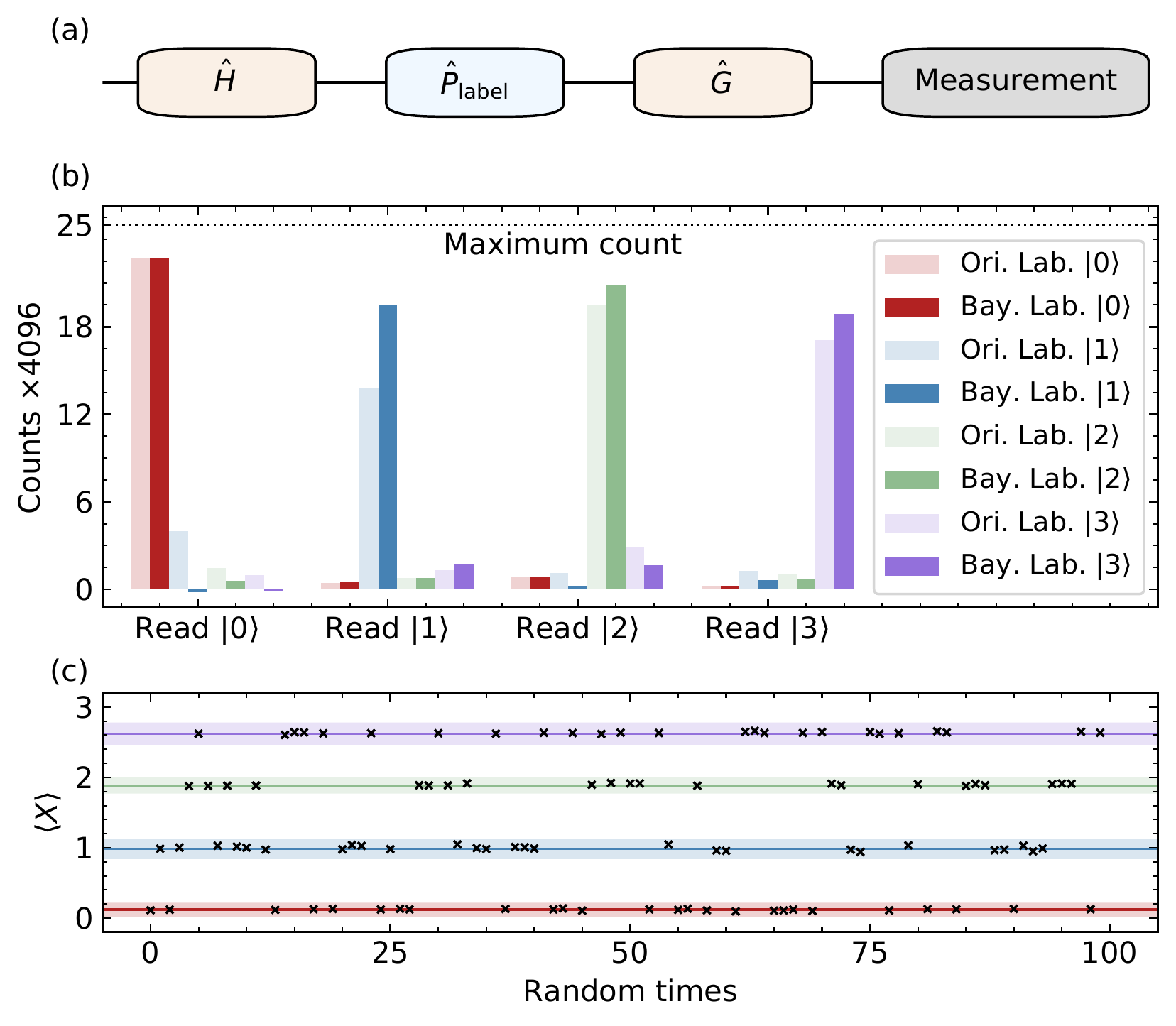}
    \caption{\label{fig:grover}
        Grover's algorithm.
        (a) The circuit with qudit initialized by Hadamard operation into an equal superposition state,
        followed by an oracle that labels one of the basis states, 
        and a standard $\hat G$ operation applied with measurement in the end.
        (b) Training for labeled state and readout results.
        We label $|0\rangle$, 
        $|1\rangle$, 
        $|2\rangle$,
        and $|3\rangle$ in the Grover circuit and collect all measurement results,
        shown in different colors respectively.
        For each labeled state,
        $25\times4096$ runs are tested and displayed in the form of a histogram
        and the light colors show original counts while dark colors show counts after Bayesian correction.
        Dotted line indicates the maximum count,
        which is the theoretical value without any error channel.
        (c) A simple test with random labeled state.
        For each test,
        the expectation $\langle X\rangle$ is calculated from $4096$ runs.
        They are remarkably differentiated from each other and the shadow shows the standard deviation trained from (b) with corresponding color. 
    }
\end{figure}

Next,
we discuss the implementation of Grover's algorithm \textcolor{black}{\cite{grover1997quantum, toyama2013quantum}} in a four-level quantum system,
where each level represents an item in an unsorted database.
The goal is to locate a specific item labeled by an oracle operation,
a black-box operation that has nontrivial effect only on the labeled level.
Without loss of generality,
we choose the oracle operation $\hat P_{\rm{label}}$ such that the amplitude of the labeled level $|\rm{label}\rangle$ flips its sign or acquires an $e^{i\pi}$ phase.

The first four levels of our transmon qudit are used,
and the operation $\hat H$ for initialization carries out $\hat H |0\rangle=1/2\sum_{n=0}^3|n\rangle$ with the matrix representation $\rm H$ taking the form
\begin{equation}
    \rm H=\frac12\begin{pmatrix}
        1& 1& 1& 1\\
        1&-1& 1&-1\\
        1& 1&-1&-1\\
        1&-1&-1& 1
    \end{pmatrix},
\end{equation}
which is the same as the generalized Hadamard operator.
It turns into an equal superposition of all basis states and with the oracle marking the specific state for Grover's algorithm to find through a generalized phase gate $\hat P_{\rm{label}}(\vec\Phi)$,
satisfying
\begin{equation}
    \phi_j=\begin{cases}
        -1,&j=\rm{label},\\
        1,&{\rm{otherwise}},
    \end{cases}.
\end{equation}
The simplest search operator $\hat G$ takes the form
\begin{equation}
    \rm{G}=\frac12\begin{pmatrix}
        -1& 1& 1& 1\\
         1&-1& 1& 1\\
         1& 1&-1& 1\\
         1& 1& 1&-1
    \end{pmatrix}.
\end{equation}
For most instances,
the search operator would be more complicated,
because an arbitrary phase rotation satisfying the phase matching requirement is needed to achieve accurate search \cite{long2001grover}.
But for $d=4$,
this phase is just $\pi$ and therefore $\hat G$ takes the original form.
At the end of the search operation,
we measure the qudit to affirm the state.
Figure~\ref{fig:grover}(a) shows the complete gate sequence.

The labeled state is read with data shown in Fig.~\ref{fig:grover}(b) in red, 
blue,
green,
and purple colors for $|0\rangle$, 
$|1\rangle$, 
$|2\rangle$,
or $|3\rangle$ being labeled respectively.
From these results,
we obtain a trained data set,
$\langle X_n\rangle=\sum_m mP_m(n)$
where $P_m(n)$ denotes the probability of reading out $|m\rangle$ in $4096$ runs when we label state $|n\rangle$,
with a standard deviation $\Delta\langle X_n\rangle$ from $25$ repetitions.
$\langle X_n\rangle\pm\Delta\langle X_n\rangle, n=0, 1, 2, 3$,
provides identification confidence interval.
100 random test results are plotted as black crosses in Fig.~\ref{fig:grover}(c),
which fits well with the range from training and confirms accurate manipulation in qudit and the reliable execution of the decomposed sequence for universal SU($d$) operations.

\subsection{Variational Quantum Eigensolver in Quantum Chemistry}

\begin{figure*}
    \includegraphics[width=7in]{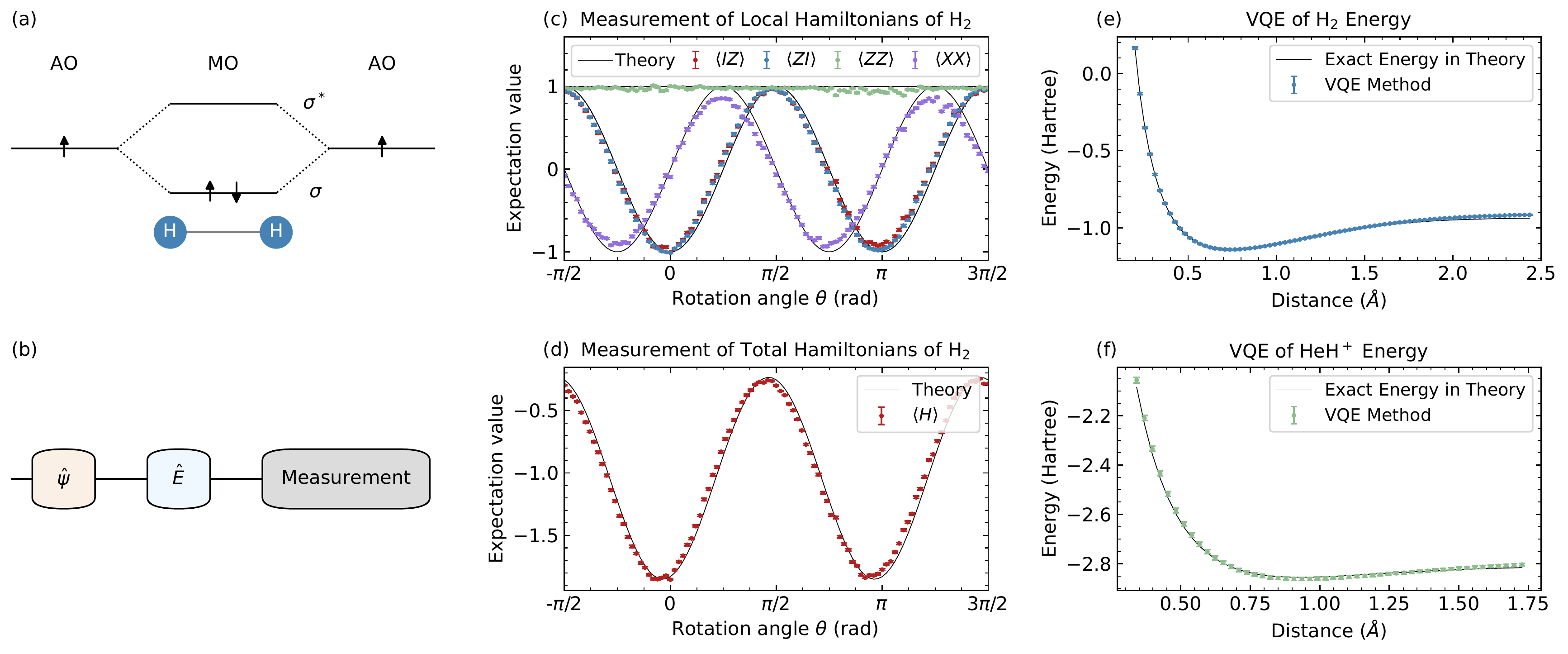}
    \caption{\label{fig:qc}
        Solving for $\rm{H}_2$ and $\rm{HeH}^+$ ground state energies with VQE.
        (a) Formation of molecular orbitals for $\rm H_2$.
        (b) VQE circuit in the qudit,
        with a trial initial wave function $\hat\Psi$ and a rotation of measurement operator $\hat E$.
        (c) Expectation for each measurement operator for local Hamiltonian in $\rm{H}_2$.
        (d) Hamiltonian of $\rm{H}_2$ as a function of rotation angle $\theta$,
        the variation parameter of trial wave function.
        (e) Energy of the ground state for $\rm{H}_2$.
        (f) Energy of the ground state for $\rm{HeH}^+$.
    }
\end{figure*}

Quantum chemistry calculation or quantum chemistry simulation is a popular application for quantum information processing.
In the noisy intermediate-scale quantum (NISQ) era,
quantum algorithm like variational quantum eigensolver (VQE) attracts a lot of attention \textcolor{black}{\cite{preskill2018quantumcomputingin, peruzzo2014variational}},
and several experiments have implemented VQE in qubit systems,
solving for ground state $\rm{H}_2$ energy in a superconducting system \cite{malley2016scalable} and $\rm{HeH}^+$ energy in a trapped-ion system \cite{shen2017quantum}.
The dimension of Hilbert space in these two early studies are limited to two qubits or four-level systems,
which we could implement (perhaps with higher qualities) in the transmon qudit.

As detailed in Appendix~\ref{app:qc},
to obtain the ground state energy of $\rm{H}_2$,
VQE minimizes the encoded Hamiltonian
\begin{equation}
    H_{\rm{H}_2}^{\rm{BK}}=a_0I+a_1IZ+a_2ZI+a_3ZZ+a_4XX,
\end{equation}
with a trial wavefunction $e^{i\theta XY}|11\rangle$,
where $X, Y,$ and $Z$ are Pauli matrices of a single qubit,
$|11\rangle$ is the Hartree-Fock state encoded by Bravyi-Kitaev transformation with Z$_2$ symmetry, and coefficients $\{a_l\}$ are calculated on classical computer.
When running in four-level system,
the two-qubit states are mapped to multi-level states,
for instance choosing $|00\rangle\to|0\rangle$,
$|01\rangle\to|1\rangle$,
$|10\rangle\to|2\rangle$,
and $|11\rangle\to|3\rangle$.
The trial wavefunction $e^{i\theta XY}|11\rangle$ is thus mapped to 
\begin{equation}
    \begin{aligned}
        e^{i\theta XY}|11\rangle&\to\hat P(\vec\Phi)\hat U^{(4)}\hat U^{(3)}\hat U^{(2)}\hat U^{(1)}\hat U^{(0)}|0\rangle,\\
        \hat U^{(0)}&=\hat U_2^{|0\rangle, |1\rangle}\left(\pi, 0, \pi, \frac\pi2\right),\\
        \hat U^{(1)}&=\hat U_2^{|1\rangle, |2\rangle}\left(\pi, 0, \pi, \frac\pi2\right),\\
        \hat U^{(2)}&=\hat U_2^{|2\rangle, |3\rangle}\left(\pi, 0, \pi, \frac\pi2\right),\\ 
        \hat U^{(3)}&=\hat U_2^{|0\rangle, |3\rangle}\left(2\theta, 0, -\pi, \frac\pi2\right),\\
        \hat U^{(4)}&=\hat U_2^{|1\rangle, |2\rangle}\left(2\theta, -\pi, 0, \frac\pi2\right),\\
        \vec\Phi&=(\pi, \pi, 0, 0)^T,\\
    \end{aligned}
\end{equation}
derived from Gaussian elimination in Algorithm~\ref{alg:gauss},
which constitutes a precise decomposition.
Measurements of $\langle IZ\rangle$,
$\langle ZI\rangle$,
$\langle ZZ\rangle$,
and $\langle XX\rangle$ follow the same experimental method as described before.
The whole process is illustrated in Fig.~\ref{fig:qc}(b).
Figures~\ref{fig:qc}(c) and \ref{fig:qc}(d) display theoretical and experimental results of local and total Hamiltonians (varying with parameters $\theta$) respectively at equilibrium point. 
Figure~\ref{fig:qc}(e) shows the experimental energy curve of the ground state of $\rm{H}_2$ as a function of internuclear distance,
which fits well with the exact value (black line).

$\rm{HeH}^+$ can also be simulated despite of being more complicated than $\rm{H}_2$ from the aspect of simulation.
The corresponding trial wavefunction is chosen as 
\begin{equation}
    \exp{\left[i\frac{\theta_1}2(IY+YI)+i\frac{\theta_2}2(XY+YX)\right]}|11\rangle,
\end{equation}
and the encoded Hamiltonian takes the form
\begin{equation}
    \begin{aligned}
        H_{\rm{HeH}^+}^{\rm{BK}}=a_0I+&a_1IZ+a_2IX+a_3ZI+a_4XI\\
                                     +&a_5ZZ+a_6ZX+a_7XZ+a_8XX.
    \end{aligned}
\end{equation}
Similar to the case of $\rm{H}_2$,
both wavefunction construction operators at given parameters and measurement operations are represented in matrix forms and decomposed by Gaussian eliminations.
The corresponding details shall not be repeated here.
Afterwards,
VQE for $\rm{HeH}^+$ is implemented on our system and the results of ground state energy are shown in Fig.~\ref{fig:qc}(f),
consistent with the exact energy obtained from theoretical calculations.
\textcolor{black}{A second figure for the energy error with fluctuations is shown in Appendix~\ref{app:qc} Fig.~\ref{fig:veqerror}.
Our experiment replaces two coupling transmon qubits with one transmon qudit,
saving crucial hardware resource.}

\section{Conclusion and Outlook\label{outlook}}

We \textcolor{black}{realize} a set of operations for readout,
calibration,
quantum state and process tomography,
and randomized benchmarking in a qudit constructed from a superconducting transmon.
We further implement three algorithms with the well-benchmarked quantum device and verify their advantages in calculation efficiency and accuracy.
\textcolor{black}{
It demonstrates the resource saving with simpler gate operations than two-qubit gates and a reduction in hardware requirement.
}

\textcolor{black}{In transmon qudit,
the couplings strength increases as $\sqrt m$,
slower than the lifetime scaling of $1/m$ for $m$ excitations, 
which consequently would erect a level limit in high-fidelity manipulations that can be achieved,
thus scalability of qudit system is to be reconsidered.}
In reference \cite{wang2020qudits},
a summary of the definitions and properties for various examples of qudit gates are provided,
including two-qudit SWAP gate,
which can be implemented by using controlled-shift gate $CX_d$ satisfying $CX_d|x\rangle|y\rangle = |x\rangle|x+y\rangle$.
Within such a framework, 
it shows that universal and complete operations can be constructed in principle.
We are hopeful with the achievement reported here,
one can go one step further by implementing two-qudit gates in the future as cross-resonant gates or some other schemes.
Meanwhile,
advanced calculation schemes,
more and improved algorithms,
and quantum simulation theories based on qudit systems are urgently needed.

On the side of hardware implementation for qudit systems,
transmons,
remain a popular choice due to their inherent multi-level structure,
which proves convenient for realizing multi-qudit quantum devices with slight changes in design.
\textcolor{black}{What's more,}
the scalability of transmon or other choiced devices in superconducting system provides many benefits for implementing qudits,
though higher qualities of such units are needed.

\textcolor{black}{Note added.
During the preparation of this work,
we become aware of several related works on the qudit-based quantum computation,
including the study on the beat note in Ramsey interference \cite{martinez2022noise},
the two-qutrit gate based on cross-resonant gate and its application \cite{fischer2022towards},
the GST protocol for qutrit \cite{cao2022efficient},
several algorithms demonstrated on a qutrit processor \cite{roy2022realization},
the gate compilation protocol for qudit-based architecture \cite{mato2023compilation},
the comparison of gate efficiency between the qudit and qubits \cite{jankovvic2023noisy}.}

\begin{acknowledgments}
    We appreciate the helpful discussion with Yu Song.
    This work is supported by the National Natural Science Foundation of China (NSFC, 
    Grant No. 11890704),
    the Beijing Natural Science Foundation (Grant No. Z190012), NSFC (Grant No. 12104055,
    Grant No. 12004042,
    Grant No. 12104056),
    and the Key Area Research and Development Program of Guang Dong Province (Grant No. 2018B030326001).
\end{acknowledgments}

\appendix 
\setcounter{figure}{0}
\setcounter{table}{0}
\renewcommand{\thefigure}{S\arabic{figure}}
\renewcommand{\thetable}{S\arabic{table}}

\section{Proof of Exchanging Generalized Phase Gate}\label{app:proof}

The proof of Eq.(~\ref{pr=rp}) is shown below:
\begin{widetext}
    \begin{equation}
        \begin{aligned}
             &\hat R_{m, n}(\theta, \phi)\hat P(\vec\Phi)\\
            =&\left[\cos\frac\theta2\hat I^{m,n}-i\sin\frac\theta2\left(\cos\phi\hat\sigma_x^{m,n}+\sin\phi\hat\sigma_y^{m,n}\right)\right.\left.\frac{}{}\right]\left(\sum_{k=m,n}e^{i\phi_k}|k\rangle\langle k|\right)\oplus\left(\sum_{k\neq m,n}e^{i\phi_k}|k\rangle\langle k|\right)\\
            =&\left[\cos\frac\theta2\left(|m\rangle\langle m|+|n\rangle\langle n|\right)-i\sin\frac\theta2\left(e^{-i\phi}|m\rangle\langle n|+e^{i\phi}|n\rangle\langle m|\right)\right]\left(\sum_{k=m,n}e^{i\phi_k}|k\rangle\langle k|\right)\oplus\left(\sum_{k\neq m,n}e^{i\phi_k}|k\rangle\langle k|\right)\\
            =&\left(\sum_{k\neq m,n}e^{i\phi_k}|k\rangle\langle k|\right)\oplus\left\{\cos\frac\theta2\left(e^{i\phi_m}|m\rangle\langle m|+e^{i\phi_n}|n\rangle\langle n|\right)-i\sin\frac\theta2\left[e^{-i(\phi-\phi_n)}|m\rangle\langle n|+e^{i(\phi+\phi_m)}|n\rangle\langle m|\right]\right\}\\       
            =&\left(\sum_{k\neq m,n}e^{i\phi_k}|k\rangle\langle k|\right)\oplus\left(\sum_{k=m,n}e^{i\phi_k}|k\rangle\langle k|\right)\left[\cos\frac\theta2\left(|m\rangle\langle m|+|n\rangle\langle n|\right)-i\sin\frac\theta2\left(e^{-i\phi'}|m\rangle\langle n|+e^{i\phi'}|n\rangle\langle m|\right)\right]\\
            =&\hat P(\vec\Phi)\hat R_{m, n}(\theta, \phi+\phi_m-\phi_n),\\
        \end{aligned}
    \end{equation}
    \end{widetext}
with $\phi'=\phi+\phi_m-\phi_n$.
Therefore it is valid to swap a generalized phase gate with any arbitrary transitions,
leading to zero duration of generalized phase gate,
as in the virtual Z strategy \cite{mckay2017efficient} discussed before.

\section{Algorithm for Gaussian Elimination}\label{app:gauss}

\begin{algorithm}
    \caption{Normal Gaussian Elimination Decomposition of SU($d$)} \label{alg:gauss}
    \SetKwFunction{func}{GED}
    \SetKwProg{Fn}{Function}{:}{}
    \Fn{\func{$d, U$}}{
        \KwIn{
            current dimension $d$ and the $d\times d$ unitary representation matrix $U$
        }
        \KwOut{
        sequences $Ans$ of SU(2) operations
        }
        \BlankLine
        initialize $Ans$, set empty\;
        \For{$j\leftarrow 1$ \KwTo $d-1$}{
            \If{not $U[d-j, d]=0$}{
                calculate a SU(2) operation $u'$ according to $U[d, d]$ and $U[d-j, d]$\;
                add answer $Ans \leftarrow Ans\cup u'$\;
                update $U\leftarrow u'U$\;
            }
        }
        add a generalized phase gate $Ans \leftarrow Ans\cup p$ making $U[d, d]$ equal to $1$\;
        \eIf{$d>1$}{
            \KwRet{$Ans\cup$\func{$d-1, U[1:d-1, 1:d-1]$}}\;
        }{
            \KwRet{$Ans$}\;
        }
    }
\end{algorithm}

Based on the Gaussian elimination,
an arbitrary SU($d$) operation can be decomposed into a sequence of SU(2) operations \cite{nielsen2010, oleary2006parallelism}.
This algorithm takes the unitary $d$-dimensional matrix representation of the SU($d$) operators as input,
and outputs a SU(2) operation sequence with generalized phase gates,
or simply expressed as the following
\begin{equation}
	\hat U_d^{0, 1, \cdots, d-1}=\hat P(\vec\Phi)\hat U_2^{m_k, n_k}\cdots\hat U_2^{m_1, n_1}\hat U_2^{m_0, n_0},
\end{equation}
where a total of $(k+1)$ SU(2) operations are required in a given order and $m_l, n_l\in\{0, 1, \cdots, d-1\}, l=0, 1, \cdots, k$.
The strategy of Gaussian elimination affects the implementation of sequences,
and the normal and the bubbling ones are described in Algorithm~\ref{alg:gauss} and Algorithm~\ref{alg:gauss2},
respectively.
Although both of them provide no more than $d(d-1)/2$ SU(2) operations,
the level structure of each SU(2) operation finally affects the pulse number required in the SU($d$) operation.
Obviously,
the normal Gaussian elimination is more suitable for the trapped ion system in Ref. \cite{Ringbauer2022},
while the bubbling one is more suitable for the qudit in this work,
which is close to the theoretical limit.

\begin{algorithm}
    \caption{Bubbling Gaussian Elimination Decomposition of SU($d$)} \label{alg:gauss2}
    \SetKwFunction{func}{GED}
    \SetKwProg{Fn}{Function}{:}{}
    \Fn{\func{$d, U$}}{
        \KwIn{
            current dimension $d$ and the $d\times d$ unitary representation matrix $U$
        }
        \KwOut{
        sequences $Ans$ of SU(2) operations only spanned on adjacent levels 
        }
        \BlankLine
        initialize $Ans$, set empty\;
        \For{$j\leftarrow 1$ \KwTo $d-1$}{
            \If{not $U[j, d]=0$}{
                calculate a SU(2) operation $u'$ according to $U[j+1, d]$ and $U[j, d]$\;
                add answer $Ans \leftarrow Ans\cup u'$\;
                update $U\leftarrow u'U$\;
            }
        }
        add a generalized phase gate $Ans \leftarrow Ans\cup p$ making $U[d, d]$ equal to $1$\;
        \eIf{$d>1$}{
            \KwRet{$Ans\cup$\func{$d-1, U[1:d-1, 1:d-1]$}}\;
        }{
            \KwRet{$Ans$}\;
        }
    }
\end{algorithm}

\section{Qudit Parameters and Experimental Setup}\label{app:params}

\begin{figure*}
    \includegraphics[width=7in]{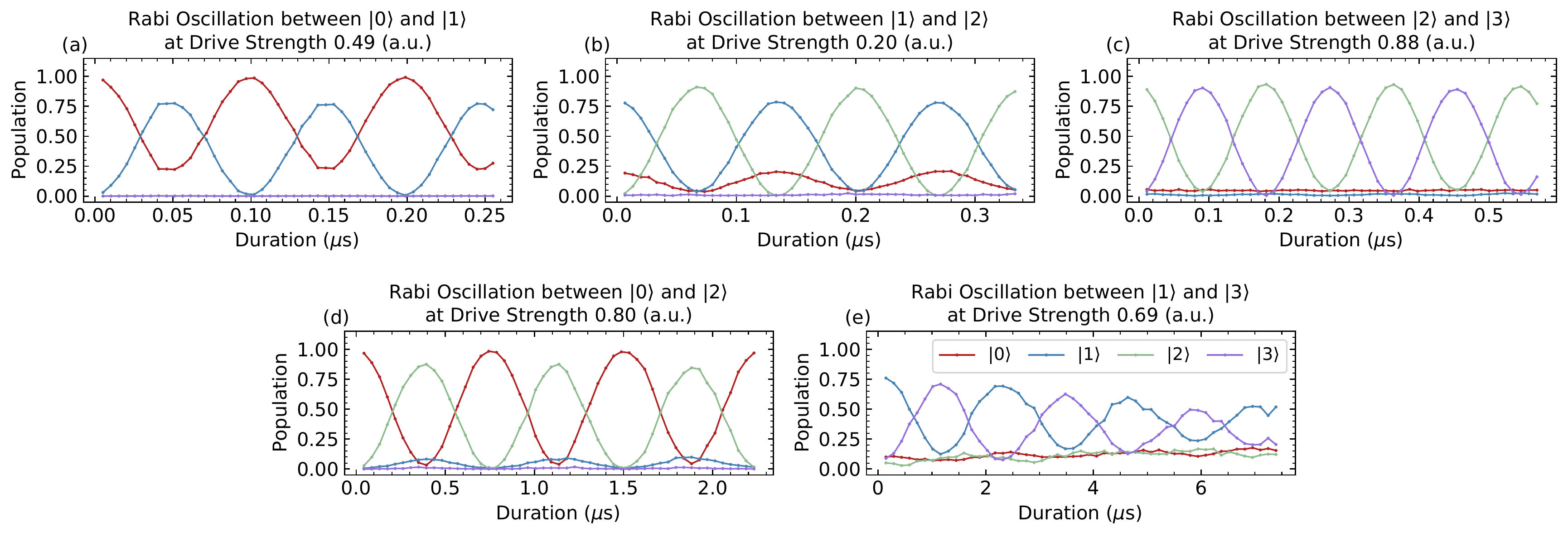}
    \caption{\label{fig:drive}
        Basic microwave-driven transitions in the transmon qudit.
        Transitions between adjacent levels are shown in (a)-(c) without readout correction or phase correction.
        Two-photon transitions are also observed and displayed in (d) and (e).
        The population damping in (e) is due to the frequency dichotomy of higher level,
        and the signal is within the first period of oscillation beat.
    }
\end{figure*}

\begin{figure}
    \includegraphics[width=3.4in]{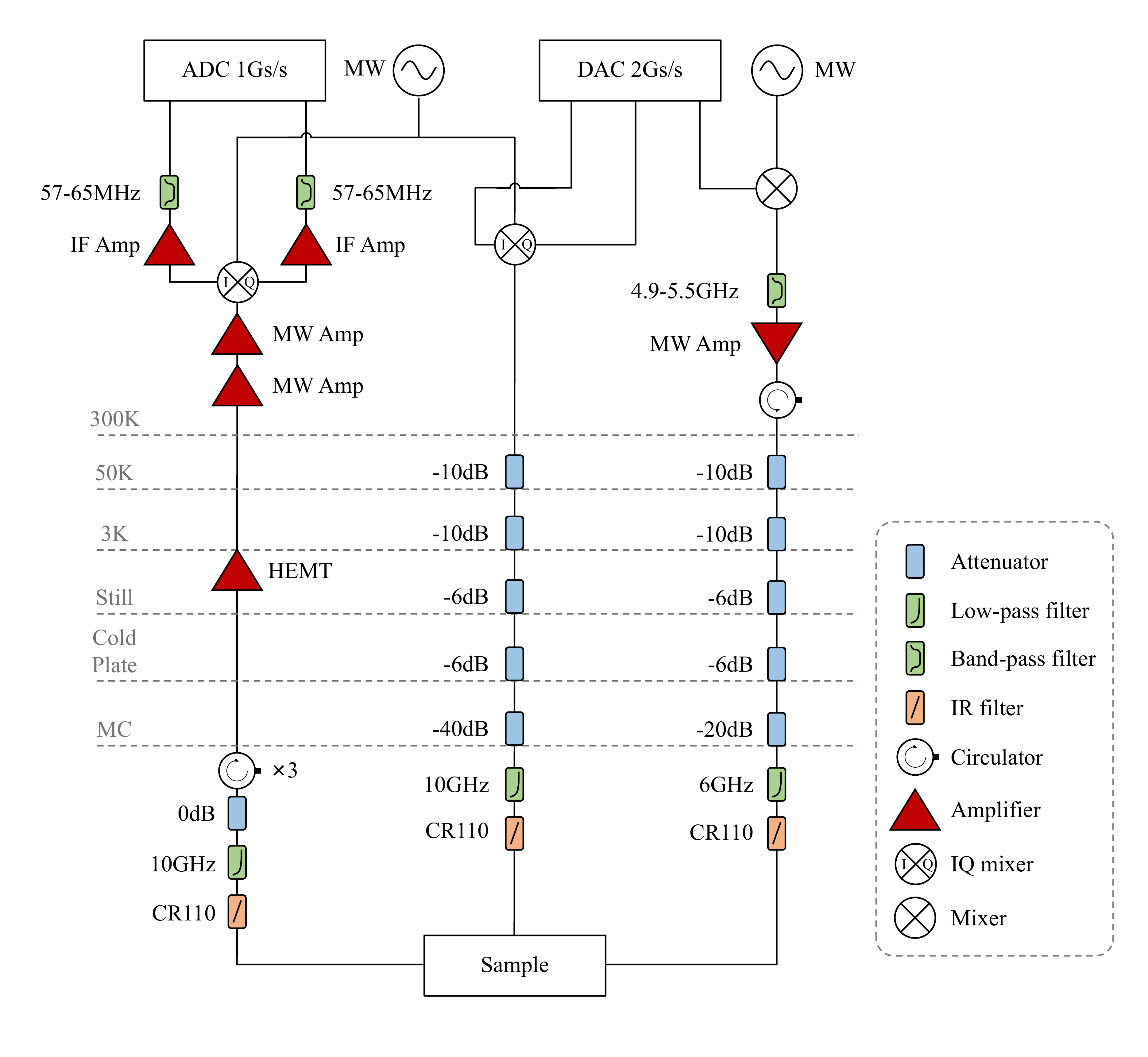}
    \caption{\label{fig:setup}
        A schematic of the measurement system includes the cryogenic and room temperature setup.
        Digital-to-analog converter (DAC) is the arbitrary waveform generator while analog-to-digital converter (ADC) is the waveform collector.
        Two microwave generators (MW) are employed to generate local signals.
    }
\end{figure}

\begin{figure*}
    \includegraphics[width=7in]{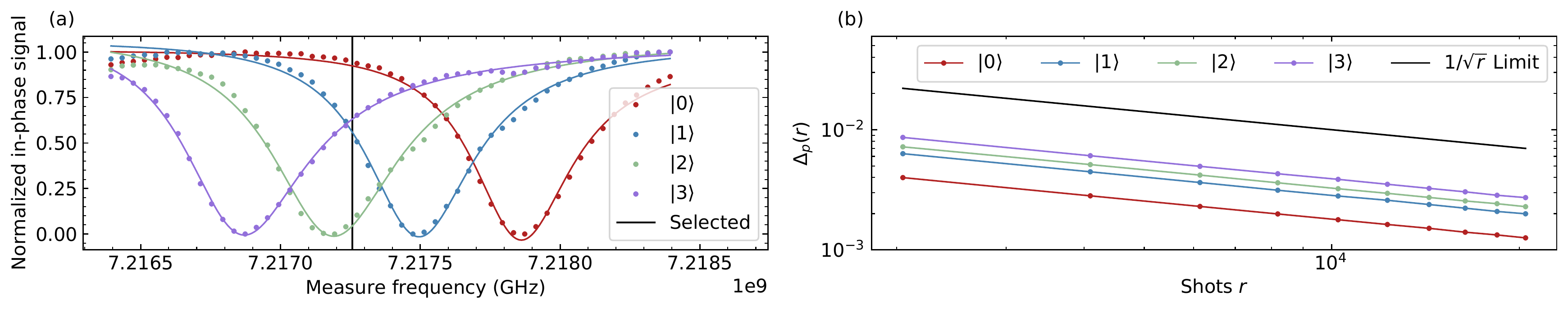}
    \caption{\label{fig:readout_std}
        \textcolor{black}{
        (a) In-phase response signal at changing measurement frequency.
        The cavity responds differently when the qubit is prepared in a different state $|k\rangle$.
        The black solid line indicates the frequency we select for state discrimination.
        (b) Standard deviation for state discrimination calculated by bootstrapping.
        From a total of $20480$ runs of $|k\rangle$ state measurements,
        $r$ runs are randomly chosen to calculate the standard deviation $\Delta_p(r)$ with $p=1$ when a run gives the measurement $|k\rangle$ and otherwise $p=0$,
        and this process is averaged over $50000$ repetitions and provides $\Delta_p(r)$ as a function of $r$,
        plotted in chosen colors for different initial states.
        The expected convergence limit $1/\sqrt{r}$ is given by the black solid line.}
    }
\end{figure*}

\begin{figure*}
    \includegraphics[width=7in]{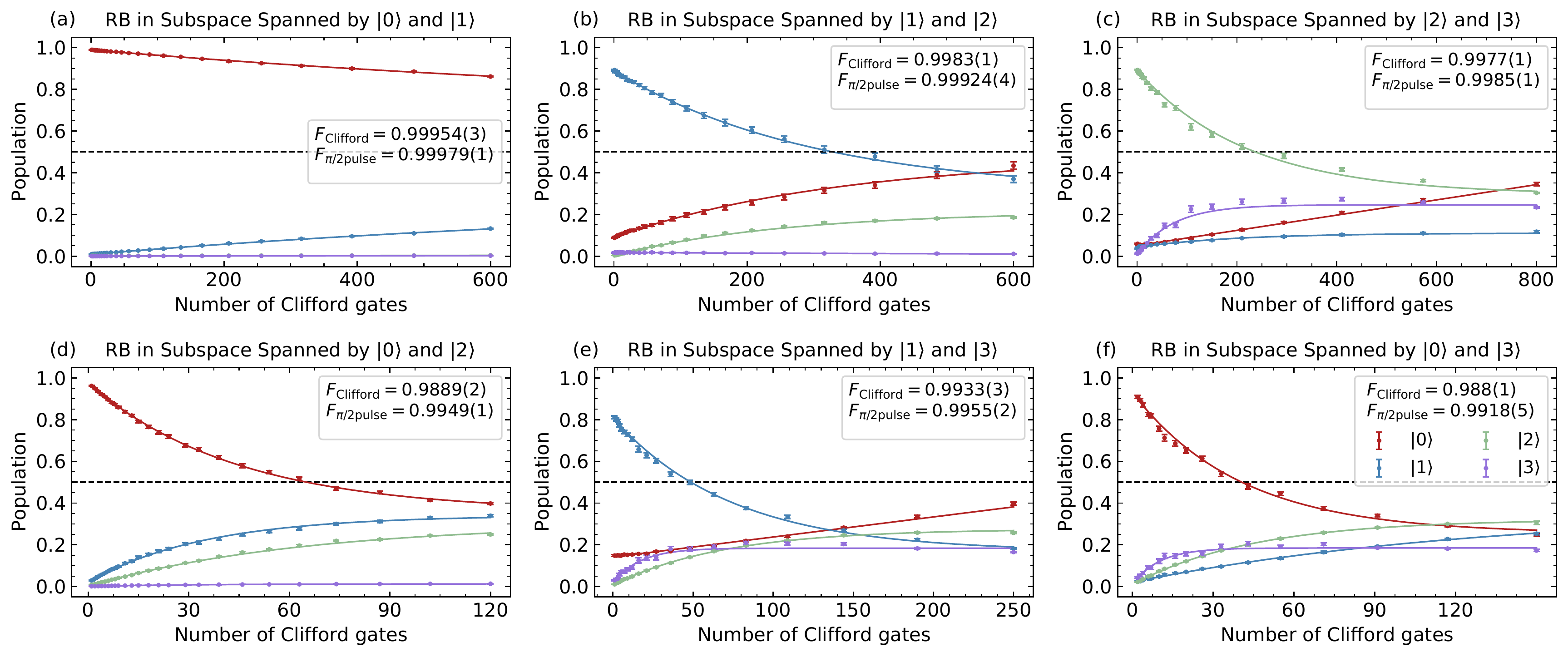}
    \caption{\label{fig:rb2}
        Randomized benchmarking on arbitrary SU(2) subspace.
		Note that in (a) $|0\rangle$ and $|1\rangle$,
        (b) $|1\rangle$ and $|2\rangle$,
        (c) $|2\rangle$ and $|3\rangle$ only adjacent transitions are used in the corresponding subspace after initial state preparation with slightly different RB sequences,
        requiring different average $\pi/2$ pulses per Clifford gate,
        numbered $2.167$,
        $2.167$,
        and $1.5$ respectively.
        (d) $|0\rangle$ and $|2\rangle$,
        (e) $|1\rangle$ and $|3\rangle$,
        (f) $|0\rangle$ and $|3\rangle$ focus on operations over the corresponding subspaces and therefore are combinations of adjacent transitions.
        All four states are measured and results are plotted as crosses with error bars and full line fitting in corresponding colors.
        Black dotted lines indicate the population of a mixed state of SU(2) after evolving for a sufficiently long time
    }
\end{figure*}

\begin{table*}
    \centering
    \caption{\label{tab:para}
        Control and Coherent Parameters of the Qudit Device.
    }
    \begin{threeparttable}
    \begin{tabular}{cc|cc|cc|cc}
        \toprule
        \makecell{adjacent one-photon \\transition frequency}& (in GHz)&\makecell{two-photon\\transition frequency}& (in GHz)&\makecell{life time of\\excited levels}& (in $\mu$s)&\makecell{dephasing time between\\labeled states}& (in $\mu$s)\\
        \hline
            $\omega_{01}/2\pi$& $5.355$&
            $\omega_{02}/2\pi$& $5.241$&
            $T^{|1\rangle}_1$& $180\pm3$&
            $T^{|0\rangle, |1\rangle}_{2, \rm{Ramsey}}$& $76\pm2$
            \\
            $\omega_{12}/2\pi$& $5.127$&
            $\omega_{13}/2\pi$& $5.000$&
            $T^{|2\rangle}_1$& $101\pm1$ &
            $T^{|1\rangle, |2\rangle}_{2, \rm{Ramsey}}$& $37\pm3$
            \\
            $\omega_{23}/2\pi$& $4.873$&
            $\bar\omega_{24}/2\pi$\tnote{*}& $4.727$&
            $T^{|3\rangle}_1$& $73\pm1$ 
            &
            $T^{|2\rangle, |3\rangle}_{2, \rm{Ramsey}}$& $22.8\pm0.3$
            \\
            $\bar\omega_{34}/2\pi$\tnote{*}& $4.581$&
            &&
            &&
            $T^{|3\rangle, |4\rangle}_{2, \rm{Ramsey}}$& $\ge 10$\\
        \bottomrule
    \end{tabular}
    \begin{tablenotes}
        \item{*} \textcolor{black}{Average transition frequency between the corresponding levels.}
    \end{tablenotes}
    \end{threeparttable}
\end{table*}

\begin{table*}
    \centering
    \caption{\label{tab:readoutpara}
        Readout Parameters of the Qudit Device.
    }
    \begin{tabular}{cc|cc|cc}
        \toprule
        \makecell{cavity response frequency \\for different qudit state $|k\rangle$} & (in GHz)&\makecell{half high\\ and half wide}& (in MHz)&other parameters& value\\
        \hline
            $\omega_{|0\rangle}/2\pi$& 7.21785& $\gamma_{|0\rangle}/2\pi$& 0.22&
            selected readout frequency $\omega_r/2\pi$& $7.21726$ GHz\\
            $\omega_{|1\rangle}/2\pi$& 7.21748& $\gamma_{|1\rangle}/2\pi$& 0.23&
            dispersive shift $\chi/2\pi$& $0.34$ MHz\\
            $\omega_{|2\rangle}/2\pi$& 7.21717& $\gamma_{|2\rangle}/2\pi$& 0.28&
            readout coupling strength $g/2\pi$& $76$ MHz\\
            $\omega_{|3\rangle}/2\pi$& 7.21683& $\gamma_{|3\rangle}/2\pi$& 0.27&&\\
        \bottomrule
    \end{tabular}
\end{table*}

Our qudit system is implemented on a transmon,
regarded as a multi-level system instead of the nominal two level qubit.
\textcolor{black}{The fabrication of our device follows the same procedure as described in Ref. \cite{wang2022towards}.}
Table \ref{tab:para} shows its \textcolor{black}{control and coherence} parameters.
$\omega_{m,n}/2\pi$ denotes the transition frequency between levels $|m\rangle$ and $|n\rangle$,
and $T^{|m\rangle}_1$ indicates the characteristic time of energy relaxation if the qudit is initialized in $|m\rangle$ state.
$T^{|m\rangle, |n\rangle}_{2, \rm{Ramsey}}$ denotes dephasing time measured via Ramsey interferometry protocol between levels $|m\rangle$ and $|n\rangle$.
Higher levels are influenced more than lower ones by charge parity,
causing fitting of the experimental data to be more difficult,
a rough estimate of dephasing is therefore given for higher levels.
Figure~\ref{fig:drive} shows results of time-dependent Rabi oscillations under single- or two-photon transition without readout correction or phase correction.
It is obvious that transitions between adjacent levels are faster than nonadjacent ones.
Therefore,
single-photon transitions are chosen as basic transitions to achieve high-fidelity manipulations.

\begin{table}
    \centering
    \caption{\label{tab:readouterror}
        An instance for preparation and readout fidelity
    }
    \begin{tabular}{c|cccc}
        \toprule
        &detect $|0\rangle$&detect $|1\rangle$& detect $|2\rangle$ & detect $|3\rangle$\\
        \hline
            prepare $|0\rangle$& 0.99104& 0.00831& 0.00060& 0.00005\\ 
            prepare $|1\rangle$& 0.05353& 0.94459& 0.00071& 0.00117\\
            prepare $|2\rangle$& 0.02428& 0.02995& 0.94498& 0.00079\\
            prepare $|3\rangle$& 0.03780& 0.00689& 0.04419& 0.91112\\
        \bottomrule
    \end{tabular}
\end{table}

The qudit device is installed in a dilution refrigerator \textcolor{black}{with details shown in Fig.~\ref{fig:setup}.}
Two microwave lines coupled to the Josephson junction and readout resonator are used to drive and detect.
Readout signals are generated from the arbitrary wave generator and up-converted to $7.217\text{ GHz}$ to probe the four levels simultaneously at room temperature.
\textcolor{black}{Table \ref{tab:readoutpara} and figure~\ref{fig:readout_std}(a) present readout response parameters.}
A statistic analysis for the readout signal is shown in Fig.~\ref{fig:readout_std}(b).
We prepare the state $|k\rangle$ and calculate the standard deviation of acquiring the measurement result $|k\rangle$ as a function of repeated runs $r$ \textcolor{black}{by bootstrapping}.
The standard deviation of experimental results decreases as $r$ increases,
\textcolor{black}{reaching the theoretical convergence rate $1/\sqrt{r}$}.
\textcolor{black}{An instance of readout fidelity matrix is shown in Tab.~\ref{tab:readouterror}.}

The drive signal is generated from the same arbitrary wave generator but in a different channel.
Different from the readout waveform,
more frequency component waves are required for qudit manipulations.
Further more,
pulses of different frequencies are combined in the same drive pulse sequence.
Phase differences between these pulses are important so that a common channel of arbitrary waveform generator is employed.
Pulse sequence is generated and up-converted to the frequencies we want via an IQ mixer at a local frequency of $4.8\text{ GHz}$.
We use $\pi/2$ cosine pulses with arbitrary phase as basic components,
whose duration between $|0\rangle$ and $|1\rangle$($|1\rangle$ and $|2\rangle$, 
$|2\rangle$ and $|3\rangle$) is $45\text{ ns}$ ($35\text{ ns}$, 
$85\text{ ns}$) with a $10\text{ ns}$ buffer.

To calibrate the $\pi/2$ pulse,
we determine drive frequency using Ramsey interferometry.
If interference with a beat is found in the Ramsey signal,
the mean of the two frequencies is taken.
Then we determine the frequency shift according to derivative reduction by adiabatic gate (DRAG) \cite{gambetta2011analytic} method for phase error.
In fact this dichotomy of transition frequency influences our manipulation capability and makes the phase error difficult to calibrate.
The amplitude of pulse would be influenced at a nonzero frequency shift,
so the above two steps are repeated for more precise calibration.
Figures \ref{fig:rb2} (a)-(c) show randomized benchmarking sequences on SU(2) subspace $|0\rangle$ and $|1\rangle$,
$|1\rangle$ and $|2\rangle$,
and $|2\rangle$ and $|3\rangle$ respectively,
with a fidelity $(99.979\pm0.001)\%$,
$(99.924\pm0.004)\%$,
and $(99.85\pm0.01)\%$ per $\pi/2$ pulse.
Rather than subspaces spanned by adjacent levels,
randomized benchmarking sequences are also applied via composite rotations from SU(2) transitions.
Figure \ref{fig:rb2} (e) shows that fidelities of $\pi/2$ pulses between $|0\rangle$ and $|2\rangle$,
$|1\rangle$ and $|3\rangle$,
and $|0\rangle$ and $|3\rangle$ are respectively $(99.49\pm0.01)\%$,
$(99.55\pm0.02)\%$,
and $(99.14\pm0.01)\%$.
As measurements on the four levels are implemented simultaneously,
leakages,
induced by energy relaxation or frequency bandwidth,
contribute to incoherent errors for SU(2) manipulations,
which may not be observed directly by randomized benchmarking (RB) in two-level systems.
Derivative reduction by adiabatic gate is usually adopted to suppress unwanted frequency components with driving pulses in qubit manipulations \cite{gambetta2011analytic}.
Inspired by the above,
a similar approach could be used to suppress unwanted frequency components in qudit manipulation.
However,
the driving scheme would become more involved,
because there are now more frequency components to suppress.

Unfortunately,
randomized benchmarking could not estimate all error messages,
due to the averaging effect of RB sequence.
Some specific types of errors, 
such as the non-Markovian error in operation,
could not be observed,
and only limited manipulation errors can be perceived.
In qubit systems,
such errors can be estimated by gate set tomography \cite{blume-kohout2017demonstration, greenbaum2015introduction}.
However,
for a qudit system related theoretical tools are still being developed or discovered.
It calls for increased theoretical efforts as analyzing the error source is of great importance to improve manipulation.

\section{Quantum State Tomography of Qudit}\label{app:qst}

Quantum state tomography (QST) is a common method to determine the density matrix of a quantum state.
QST of qudit system was presented in nuclear-magnetic-resonance system \cite{bonk2004quantum},
and we follow the same method.
To measure elements of an arbitrary density matrix of qudit, 
for example,
a four-level system,
we construct 12 operations $\hat M_l, l=0, 1, \cdots, 11$,
as in the following,
\begin{equation}
    \begin{aligned}
        \hat M_0&=\hat R_{0, 1}\left(-\frac\pi2, 0\right),&\\
        \hat M_1&=\hat R_{1, 2}\left(-\frac\pi2, 0\right),&\\
        \hat M_2&=\hat R_{2, 3}\left(-\frac\pi2, 0\right),&\\
        \hat M_3&=\hat R_{0, 1}\left(-\frac\pi2, \frac\pi2\right),&\\
        \hat M_4&=\hat R_{1, 2}\left(-\frac\pi2, \frac\pi2\right),&\\
        \hat M_5&=\hat R_{2, 3}\left(-\frac\pi2, \frac\pi2\right),&\\
        \hat M_6&=\hat R_{1, 2}\left(-\frac\pi2, 0\right)\hat R_{0, 1}\left(-\frac\pi2, 0\right),&\\
        \hat M_7&=\hat R_{1, 2}\left(-\frac\pi2, \frac\pi2\right)\hat R_{0, 1}\left(-\frac\pi2, 0\right),&\\
        \hat M_8&=\hat R_{2, 3}\left(-\frac\pi2, 0\right)\hat R_{1, 2}\left(-\frac\pi2, 0\right),&\\
        \hat M_9&=\hat R_{2, 3}\left(-\frac\pi2, \frac\pi2\right)\hat R_{1, 2}\left(-\frac\pi2, 0\right),&\\
        \hat M_{10}&=\hat R_{2, 3}\left(-\frac\pi2, \frac\pi2\right)\hat R_{1, 2}\left(-\frac\pi2, \frac\pi2\right)\hat R_{0, 1}\left(-\frac\pi2, \frac\pi2\right),&\\
        \hat M_{11}&=\hat R_{2, 3}\left(-\frac\pi2, 0\right)\hat R_{1, 2}\left(-\frac\pi2, 0\right)\hat R_{1, 2}\left(-\frac\pi2, 0\right).&\\
    \end{aligned}
\end{equation}
For an unknown quantum state $\hat\rho$,
we apply each operation after state preparation,
followed by a measurement gate which projects the state into one of the four eigenstates $|0\rangle$,
$|1\rangle$,
$|2\rangle$,
and $|3\rangle$,
the respective probabilities $P_{l, k}=\langle k|\hat M_l^\dagger\hat\rho\hat M_l|k\rangle$ with $l=0, 1, \cdots, 11$ and $k=0, 1, 2, 3$ are obtained,
satisfying $\sum_k P_{l, k}=1$.
Elements of $\hat\rho$ are solved according to
\begin{equation}
    \begin{aligned}
        \rho_{0, 0}&=(P_{2, 0}+P_{5, 0})/2,&\\
        \rho_{1, 1}&=(P_{2, 1}+P_{5, 2})/2,&\\
        \rho_{2, 2}&=(P_{0, 2}+P_{0, 3})/2,&\\
        \rho_{3, 3}&=(P_{0, 3}+P_{3, 3})/2,&\\
        \\
        x_{0, 1}&=(P_{3, 0}-P_{3, 1})/2,&\\
        x_{1, 2}&=(P_{4, 1}-P_{4, 2})/2,&\\
        x_{2, 3}&=(P_{5, 2}-P_{5, 3})/2,&\\
        y_{0, 1}&=(P_{0, 0}-P_{0, 1})/2,&\\
        y_{1, 2}&=(P_{1, 1}-P_{1, 2})/2,&\\
        y_{2, 3}&=(P_{2, 2}-P_{2, 3})/2,&\\
        x_{0, 2}&=(P_{6, 1}-P_{6, 2}-\sqrt2y_{1, 2})/\sqrt2,&\\
        y_{0, 2}&=(P_{7, 2}-P_{7, 1}+\sqrt2x_{1, 2})/\sqrt2,&\\
        x_{1, 3}&=(P_{8, 2}-P_{8, 3}-\sqrt2y_{2, 3})/\sqrt2,&\\
        y_{1, 3}&=(P_{9, 3}-P_{9, 2}+\sqrt2x_{2, 3})/\sqrt2,&\\
        x_{0, 3}&=P_{10, 2}-P_{10, 3}-\sqrt2x_{2, 3}+x_{1, 3},&\\
        y_{0, 3}&=P_{11, 3}-P_{11, 2}+\sqrt2y_{2, 3}+x_{1, 3},&\\
        \\
        \rho_{0, 1}&=x_{0, 1}+iy_{0, 1},&\\
        \rho_{0, 2}&=x_{0, 2}+iy_{0, 2},&\\
        \rho_{0, 3}&=x_{0, 3}+iy_{0, 3},&\\
        \rho_{1, 2}&=x_{1, 2}+iy_{0, 2},&\\
        \rho_{1, 3}&=x_{1, 3}+iy_{0, 3},&\\
        \rho_{2, 3}&=x_{2, 3}+iy_{0, 3}.\label{rho}
    \end{aligned}
\end{equation}
Unfortunately, 
the state obtained this way does not make full use of the information in the measurement data.
An alternative method is to apply maximum likelihood estimation (MLE) according to properties of the density matrix \cite{myung2003tutorial, knee2018maximum},
with the simple estimation $\hat\rho$ from Eq. (\ref{rho}) regarded as initial guess to reduce the impact of other undesirable errors on the output density matrix.

\section{Quantum Process Tomography of Qudit}\label{app:qpt}

\begin{algorithm}
    \caption{Construction of generators of SU($d$) matrix}\label{alg:sun}
    \KwIn{
        order $d$ of SU($d$).
    }
    \KwOut{
        sequences $Ans$ of SU($d$) generators in matrix representation.
    }
    initialize $Ans$, set empty\;
    add answer $Ans \leftarrow Ans\cup \rm{I}_{d\times d}$\;
    \For{$j\leftarrow2$ \KwTo $d$}{
        \For{$k\leftarrow0$ \KwTo $j-2$}{
            add answer $Ans \leftarrow Ans\cup \rm{\rho}_{m, n}=\begin{cases}
                1, &m=k, n=j-1,\\
                1, &m=j-1, n=k,\\
                0, &{\rm{otherwise}},\\
            \end{cases}$\;
            add answer $Ans \leftarrow Ans\cup \rm{\rho}_{m, n}=\begin{cases}
                -i, &m=k, n=j-1,\\
                i, &m=j-1, n=k,\\
                0, &{\rm{otherwise}},\\
            \end{cases}$\;
        }
        add answer $Ans \leftarrow Ans\cup \rm{\rho}_{m, n}=\begin{cases}
            \frac1{\sqrt{j(j-1)/2}}, &m=n=0, 1, \cdots, j-2,\\
            \frac{1-j}{\sqrt{j(j-1)/2}}, &m=n=j-1,\\
            0, &{\rm{otherwise}},\\
        \end{cases}$\;
    }
    \KwRet{$Ans$}\;
\end{algorithm}

Quantum process tomography (QPT) is based on quantum state tomography \textcolor{black}{\cite{Ringbauer2022}},
providing a convenient visible method to characterize quantum process.
Like QPT of qubit,
we initialize our $d$-level system in a given state,
then apply the process we want to determine,
and end with QST to measure the final state.
In the textbook Ref.~\cite{nielsen2010},
for arbitrary state $|m\rangle\langle n|$,
which is obviously one of the basis of density matrix,
the final state after the process $\mathcal E(|m\rangle\langle n|)$ satisfies
\begin{equation}
    \begin{aligned}
    \mathcal E(|m\rangle\langle n|)=&\mathcal E(|+\rangle\langle+|)+i\mathcal E(|-\rangle\langle-|)\\
    &-\frac{1+i}2\mathcal E(|m\rangle\langle m|)-\frac{1+i}2\mathcal E(|n\rangle\langle n|),
    \end{aligned}
\end{equation} 
where $|+\rangle=(|m\rangle+|n\rangle)/\sqrt2$ and $|-\rangle=(|m\rangle+i|n\rangle)/\sqrt2$,
and pure state $|n\rangle$,
$|m\rangle$,
$|+\rangle$,
or $|-\rangle$ can be prepared from the initial state $|0\rangle$.
Our implementation,
however,
tests another set of initial states
\begin{equation}
    |a_m\rangle\equiv|m\rangle,
\end{equation}
$m=0, 1, \cdots, d-1$,
and
\begin{equation}
    \begin{aligned}
        |a_{m, n, 0}\rangle&\equiv(|m\rangle-|n\rangle)/\sqrt2,\\
        |a_{m, n, 1}\rangle&\equiv(|m\rangle-i|n\rangle)/\sqrt2,\\
        |a_{m, n, 2}\rangle&\equiv(|m\rangle+i|n\rangle)/\sqrt2,
    \end{aligned}
\end{equation}
for $m, n=0, 1, \cdots, d-1$ and $m<n$.
It is obvious the first $d$ items initialize the system via
\begin{equation}
    \hat a_m=\hat R_{m-1, m}\left(\pi, \frac\pi2\right)\hat R_{m-2, m-1}\left(\pi, \frac\pi2\right)\cdots\hat R_{0, 1}\left(\pi, \frac\pi2\right),
\end{equation}
with $\hat a_m|0\rangle=|a_m\rangle$ and therefore
\begin{equation}
    \begin{aligned}
        \hat a_{m, n, 0}&=\hat R_{m, n}\left(\frac\pi2, -\frac\pi2\right)\hat a_m,\\
        \hat a_{m, n, 1}&=\hat R_{m, n}\left(\frac\pi2, 0\right)\hat a_m,\\
        \hat a_{m, n, 2}&=\hat R_{m, n}\left(\frac\pi2, -\frac\pi2\right)\hat a_m,
    \end{aligned}
\end{equation}
with $\hat a_{m,n,k}|0\rangle=|a_{m,n,k}\rangle$ for $k=0, 1, 2$.
A total of $d+3d(d-1)/2=d(3d-1)/2$ initial states are prepared accordingly.
After the final states are measured with QST,
we have
\begin{equation}
    \begin{aligned}
        -\mathcal E(|m\rangle\langle n|)=&\mathcal E(|a_{m, n, 0}\rangle\langle a_{m, n, 0}|)+i\mathcal E(|a_{m, n, 1}\rangle\langle a_{m, n, 1}|)\\
        &-\frac{1+i}2\mathcal E(|m\rangle\langle m|)-\frac{1+i}2\mathcal E(|n\rangle\langle n|),\\
        -\mathcal E(|n\rangle\langle m|)=&\mathcal E(|a_{m, n, 0}\rangle\langle a_{m, n, 0}|)+i\mathcal E(|a_{m, n, 2}\rangle\langle a_{m, n, 2}|)\\
        &-\frac{1+i}2\mathcal E(|m\rangle\langle m|)-\frac{1+i}2\mathcal E(|n\rangle\langle n|).\\
    \end{aligned}
\end{equation}
If we regard state set $\{|0\rangle\langle0|, |0\rangle\langle1|, \cdots, |0\rangle\langle d-1|, |1\rangle\langle0|, |1\rangle\langle1|, \cdots, |1\rangle\langle d-1|, \cdots, |d-1\rangle\langle0|, |d-1\rangle\langle1|, \cdots, |d-1\rangle\langle d-1|\}$ as a group of basis,
a superoperator representation can be recovered from $\mathcal E(|m\rangle\langle n|)$,
labeled by $\hat \chi_0$ whose matrix representation is of dimension $d^2\times d^2$.
In this work,
an alternative group of basis to represent superoperator is taken,
leading to satisfactory matrix representation of the superoperator.
We choose identity matrix of order $d$,
$d^2-1$ generators of SU($d$) matrix,
and sequence $\lambda_0, \lambda_1, \cdots, \lambda_{d^2-1}$ which are obtained following from Algorithm~\ref{alg:sun},
as our basis $\{\hat\lambda_k\}$ to represent the superoperator.
The relationship,
\begin{equation}
    \hat\rho=\sum_{k,l}\hat\lambda_k\hat\rho\hat\lambda_l^\dagger\chi_{k,l},
\end{equation}
with $\hat\rho$ an arbitrary density operator of the system,
describes the calculation of $\hat\chi$,
and the results are shown in Fig.~\ref{fig:dft}(b) and (c) for $d=3$ and $d=4$.

Maximum likelihood estimation (MLE) can also be adapted to QPT.
One can choose to use MLE in QST for each step to aim for a more precise state tomography result,
or directly using it in the whole calculation according to properties of matrix representation of superoperator under the special basis chosen here.
Positive semi-definite matrix allows for choosing MLE.

For $d=3$,
\begin{equation}
    \begin{aligned}
        &\lambda_0=\begin{pmatrix}
             1& 0& 0\\
             0& 1& 0\\
             0& 0& 1
        \end{pmatrix},&
        &\lambda_1=\begin{pmatrix}
             0& 1& 0\\
             1& 0& 0\\
             0& 0& 0
        \end{pmatrix},&\\
        &\lambda_2=\begin{pmatrix}
             0&-i& 0\\
             i& 0& 0\\
             0& 0& 0
        \end{pmatrix},&
        &\lambda_3=\begin{pmatrix}
             1& 0& 0\\
             0&-1& 0\\
             0& 0& 0
        \end{pmatrix},&\\
        &\lambda_4=\begin{pmatrix}
             0& 0& 1\\
             0& 0& 0\\
             1& 0& 0
        \end{pmatrix},&
        &\lambda_5=\begin{pmatrix}
             0& 0&-i\\
             0& 0& 0\\
             i& 0& 0
        \end{pmatrix},&\\
        &\lambda_6=\begin{pmatrix}
             0& 0& 0\\
             0& 0& 1\\
             0& 1& 0
        \end{pmatrix},&
        &\lambda_7=\begin{pmatrix}
             0& 0& 0\\
             0& 0&-i\\
             0& i& 0
        \end{pmatrix},&\\
        &\lambda_8=\frac1{\sqrt3}\begin{pmatrix}
             1& 0& 0\\
             0& 1& 0\\
             0& 0&-2
       \end{pmatrix},&
    \end{aligned}
\end{equation}
and for $d=4$,
\begin{equation}
    \begin{aligned}
        &\lambda_0=\begin{pmatrix}
            1& 0& 0& 0\\
            0& 1& 0& 0\\
            0& 0& 1& 0\\
            0& 0& 0& 1
        \end{pmatrix},&
        &\lambda_1=\begin{pmatrix}
            0& 1& 0& 0\\
            1& 0& 0& 0\\
            0& 0& 0& 0\\
            0& 0& 0& 0
        \end{pmatrix},&\\
        &\lambda_2=\begin{pmatrix}
            0&-i& 0& 0\\
            i& 0& 0& 0\\
            0& 0& 0& 0\\
            0& 0& 0& 0
        \end{pmatrix},&
        &\lambda_3=\begin{pmatrix}
            1& 0& 0& 0\\
            0&-1& 0& 0\\
            0& 0& 0& 0\\
            0& 0& 0& 0
        \end{pmatrix},&\\
        &\lambda_4=\begin{pmatrix}
            0& 0& 1& 0\\
            0& 0& 0& 0\\
            1& 0& 0& 0\\
            0& 0& 0& 0
        \end{pmatrix},&
        &\lambda_5=\begin{pmatrix}
            0& 0&-i& 0\\
            0& 0& 0& 0\\
            i& 0& 0& 0\\
            0& 0& 0& 0
        \end{pmatrix},&\\
        &\lambda_6=\begin{pmatrix}
            0& 0& 0& 0\\
            0& 0& 1& 0\\
            0& 1& 0& 0\\
            0& 0& 0& 0
        \end{pmatrix},&
        &\lambda_7=\begin{pmatrix}
            0& 0& 0& 0\\
            0& 0&-i& 0\\
            0& i& 0& 0\\
            0& 0& 0& 0
        \end{pmatrix},&\\
        &\lambda_8=\frac1{\sqrt3}\begin{pmatrix}
            1& 0& 0& 0\\
            0& 1& 0& 0\\
            0& 0&-2& 0\\
            0& 0& 0& 0
        \end{pmatrix},&
        &\lambda_9=\begin{pmatrix}
            0& 0& 0& 1\\
            0& 0& 0& 0\\
            0& 0& 0& 0\\
            1& 0& 0& 0
        \end{pmatrix},&\\
        &\lambda_{10}=\begin{pmatrix}
            0& 0& 0&-i\\
            0& 0& 0& 0\\
            0& 0& 0& 0\\
            i& 0& 0& 0
        \end{pmatrix},&
        &\lambda_{11}=\begin{pmatrix}
            0& 0& 0& 0\\
            0& 0& 0& 1\\
            0& 0& 0& 0\\
            0& 1& 0& 0
        \end{pmatrix},&\\
        &\lambda_{12}=\begin{pmatrix}
            0& 0& 0& 0\\
            0& 0& 0&-i\\
            0& 0& 0& 0\\
            0& i& 0& 0
        \end{pmatrix},&
        &\lambda_{13}=\begin{pmatrix}
            0& 0& 0& 0\\
            0& 0& 0& 0\\
            0& 0& 0& 1\\
            0& 0& 1& 0
        \end{pmatrix},&\\
        &\lambda_{14}=\begin{pmatrix}
            0& 0& 0& 0\\
            0& 0& 0& 0\\
            0& 0& 0&-i\\
            0& 0& i& 0
        \end{pmatrix},&
        &\lambda_{15}=\frac1{\sqrt6}\begin{pmatrix}
            1& 0& 0& 0\\
            0& 1& 0& 0\\
            0& 0& 1& 0\\
            0& 0& 0&-3
        \end{pmatrix}.&
    \end{aligned}
\end{equation}

\section{Construction of SU(3) and SU(4) Clifford Operations}\label{app:clifford}

\begin{algorithm}
    \caption{Construction of SU($d$) Clifford operations} \label{alg:clifford}
    \SetKwFunction{func}{Clifford}
    \SetKwProg{Fn}{Function}{:}{}
    \Fn{\func{$n=1, d$}}{
        \KwIn{
            qudit number $n$ default to 1 and qudit dimension $d$
        }
        \KwOut{
            Clifford group for SU($d$)
        }
        \BlankLine
        initialize $N_{SU(d)} \leftarrow N_{P_d^n}\times N_{Sp(d)}$\;
        initialize set $Group \leftarrow [F_d,S_d,Z_d,X_d]$\;
		initialize $h\leftarrow0$, $j\leftarrow0$, $L\leftarrow len(Group)$\;
        \While{$h<L$}{
            \For{$k \leftarrow j$ \KwTo $L$}{
					$g\leftarrow Group[h]*Group[k]$\;
                    $Group\leftarrow Group \cup g$, a global complex factor allowed\;
					$g\leftarrow Group[k]*Group[h]$\;
                    $Group\leftarrow Group \cup g$, a global complex factor allowed\;
                }
            \If{$len(Group)=N_{SU(d)}$}{
                \KwRet{$Group$}\;
            }
            \eIf{$len(Group)=L$}{$h\leftarrow h+1$\;$j\leftarrow0$\;}{$j\leftarrow L$\;$L\leftarrow len(Group)$\;}
        }
    }
\end{algorithm}

For the Clifford group $\bm{C}_d^n$ and Pauli group $\bm{P}_d^n$,
where $d$ is the qudit Hilbert-space dimension and $n$ is the number of qudits.
The quotient group $\bm{C}_d^n/\bm{P}_d^n$ is isomorphic to the symplectic group $Sp(dn)$.
The number of the Clifford group elements can be calculated by $N_{P_d^n}\times N_{Sp(dn)}$ for arbitrary $\bm{C}_d^n$.
We can generate single qudit Clifford group with the generators $F_d$, 
$P_d$,
$Z_d$,
and $X_d$.
With the orthonormal computational basis ${|s\rangle;s\in \mathbb{Z}_d}$, 
$\mathbb{Z}_d:={0,1,...,d-1}$, 
$F_d|s\rangle:=\frac{1}{\sqrt{d}}\sum_{s'\in{\mathbb{Z}_d}}\omega^{ss'}|s'\rangle$ which is the quantum Fourier transform, 
and the phase gate $P|s\rangle:=\omega^{\frac{s(s+\rho_d)}{2}}|s\rangle$, 
with $\rho_d=1$ for odd $d$ and $\rho_d=0$ otherwise,
to obtain $X_d|s\rangle = |s\oplus1\rangle$ and $Z_d|s\rangle = \omega^s|s\rangle$, 
where $\omega:=\exp(2\pi i/d)$.
To construct Clifford group operations for SU($d$), 
we follow the Algorithm~\ref{alg:clifford}.

\section{Details for Cyclic Permutation Parity Check}\label{app:dft}

\subsection{Parity Check of Cyclic Permutation}\label{app:parity}

One application of $\rm{DFT}$ is to check the parity of cyclic permutation.
In this algorithm,
for a permutation of length $d$ we want to check,
the qudit is initialized into a coherent superposition state $\rm{DFT}|m\rangle$,
and then permutation operation $\hat U_k, k=1, 2, \cdots, d!$ is applied.
Before reading out the qudit,
an inverse DFT,
labeled by $\rm{DFT}^{-1}$,
is applied to transform the state.
Different outcomes from the readout reveal parities of cyclic permutations.
We remind that choice of the initial state affects result of the algorithm with $m=0$ the trivial case and can be neglected,
and $m$ and $d$ are co-prime numbers,
i.e. $\gcd(m, d)=1$.
Parities of cyclic permutations can be directly measured from populations of $|m\rangle$ and $|d-m\rangle$,
with the former for even parity and the latter for odd parity.
If other initial states are chosen,
the results would be more complicated,
but remain completely determined.
For example,
taking $\gcd(m, d)=t$, 
then $d=qt$,
the subgroup $\mathcal G(m, d)=\langle\hat Q\rangle$ of permutations of length $d$ satisfies
\begin{equation}
    \hat Q=\begin{bmatrix}j&(j+q)\mod d\\(j+q)\mod d&j\end{bmatrix},
\end{equation}
$j=0, 1, \cdots, d-1$.
If the readout of $\hat U_k$ is still $|m\rangle$,
$\hat U_k\in g\mathcal C_{d, \rm{even}}, \forall g\in\mathcal G(m, d)$,
where $C_{d, \rm{even}}$ is the set of $d$-length cyclic permutation of even parity.
Whereas if the readout gives $|d-m\rangle$,
$\hat U_k\in g\mathcal C_{d, \rm{odd}}, \forall g\in\mathcal G(m, d)$.
Obviously $t=1$ corresponds to the trivial case where $\mathcal G(m, d)$ is the identity group.

For permutations of length-$3$,
$\gcd(2, 3)=1$,
we know the parity from $|2\rangle$ or $|1\rangle$ readout at the end of a circuit.
The cyclic operations are from
\begin{equation}
    \hat U_1=\begin{bmatrix}0&1&2\\0&1&2\end{bmatrix},
    \hat U_4=\begin{bmatrix}0&1&2\\1&2&0\end{bmatrix},
    \hat U_5=\begin{bmatrix}0&1&2\\2&0&1\end{bmatrix},
\end{equation}
and implicate even parity for final state $|2\rangle$,
while
\begin{equation}
    \hat U_2=\begin{bmatrix}0&1&2\\0&2&1\end{bmatrix},
    \hat U_3=\begin{bmatrix}0&1&2\\1&0&2\end{bmatrix},
    \hat U_6=\begin{bmatrix}0&1&2\\2&1&0\end{bmatrix},
\end{equation}
give odd parity for final state $|1\rangle$.
However for permutations of length-$4$,
a special case satisfying $N=2m$,
both parities give the same readout and 
\begin{equation}
    \bigcup_{g\in\mathcal G(2, 4)} g C(4)=C(4),
\end{equation}
where $C(4)$ is the set of all length-$4$ cyclic permutations and $|C(4)|=8$.
In other words,
parity check of length-$4$ cyclic permutations is invalid but cyclic permutations can still be distinguished with
\begin{equation}
    \begin{aligned}
        \hat U_1   &=\begin{bmatrix}0&1&2&3\\0&1&2&3\end{bmatrix},
        \hat U_6   &=\begin{bmatrix}0&1&2&3\\0&3&2&1\end{bmatrix},\\
        \hat U_8   &=\begin{bmatrix}0&1&2&3\\1&0&3&2\end{bmatrix},
        \hat U_{10}&=\begin{bmatrix}0&1&2&3\\1&2&3&0\end{bmatrix},\\
        \hat U_{15}&=\begin{bmatrix}0&1&2&3\\2&1&0&3\end{bmatrix},
        \hat U_{17}&=\begin{bmatrix}0&1&2&3\\2&3&0&1\end{bmatrix},\\
        \hat U_{19}&=\begin{bmatrix}0&1&2&3\\3&0&1&2\end{bmatrix},
        \hat U_{24}&=\begin{bmatrix}0&1&2&3\\3&2&1&0\end{bmatrix},\\
    \end{aligned}
\end{equation}
from experiment results.

\subsection{The Simplest Construction of Permutation Operations}\label{app:uk}

\begin{algorithm}
    \caption{Bubble Construction of $\hat U_k$} \label{alg:uk}
    \KwIn{
        an array $\{p_{k, j}\}, j=0, 1, \cdots, d-1$ representing $U_k$ to be constructed.
    }
    \KwOut{
        sequences $Ans$ of SU(2) operations in adjacent levels.
    }
    initialize $Ans$, set empty\;
    \While{True}{
        initialize flag $f\leftarrow True$\;
        \For{$j\leftarrow 0$ \KwTo $d-2$}{
            \If{$p_{k, j}>p_{k, j+1}$}{
                swap $p_{k, j}$ and $p_{k, j+1}$\;
                add an adjacent transposition $Ans \leftarrow Ans\cup\hat X_{j, j+1}$\;
                update flag $f\leftarrow False$\;
            }
        }
        \If{f}{
            \KwRet{$Ans$}\;
        }
    }
\end{algorithm}

In order to construct $\hat U_k$ using pulses as short as possible,
some considerations can help to simplify pulse sequence of $\hat U_k$ according to the nature of permutation and commutation,
inspired by bubble sort algorithm from Ref. \cite{cormen2010}.
For a permutation $\hat U_k$ of length $d$,
\begin{equation}
    \hat U_k=\begin{bmatrix}
        0&1&\cdots&N-1\\
        p_{k, 0}&p_{k, 1}&\cdots&p_{k, d-1}
    \end{bmatrix},
\end{equation}
satisfying $\hat U_k|j\rangle=|p_{k, j}\rangle$,
where $p_{k, j}\in\{0, 1, \cdots, d-1\}, \forall j\in\{0, 1, \cdots, d-1\}$ and $\forall j_1\neq j_2, p_{k, j_1}\neq p_{k, j_2}$,
and $k$ indicates ascending lexicographical order of $p_{k, 1}, p_{k, 2}, \cdots, p_{k, d}$.
Therefore each $\hat U_k$ takes its matrix representation $\mathbf M(\hat U_k)$ and it is easy to prove that $\hat U_k$ is unitary.
While normal Gaussian elimination is certainly a universal method to translate matrix representations into pulse sequences,
it is a bit awkward in this situation,
with each permutation equivalent to product of several transpositions.
The basic pulse $\hat R_{j, j+1}(\theta, \phi)$ we employ actually corresponds to transitions between adjacent levels,
and using only adjacent transpositions can construct arbitrary permutations,
and finding minimum numbers of adjacent transpositions is of great importance to pulse sequence simplification.
A transposition between adjacent levels can be written as $\hat X_{j, j+1}=|j+1\rangle\langle j|+|j\rangle\langle j+1|$.

A simple perspective to understand the lower bound of adjacent transposition numbers comes from inversion pairs of a permutation.
For a given permutation,
inversion pairs are determined and an adjacent transposition would increase or decrease an inversion.
Number of adjacent transpositions is no less than inverse pairs in a permutation.
If a sequence with number of adjacent transpositions equal to inverse pairs,
the minimal decomposition is found.

We can use bubbling Gaussian elimination to generate pulse sequence of $\hat U_k$,
with the same results obtained.
Hopefully,
this helps to understand why the bubbling elimination corresponds to the limit of theoretical complexity of the transmon qudit.

\section{VQE in Qudit} \label{app:qc}

\subsection{Framework of VQE}

For a molecule consisting of $N$ nuclei and $M$ (valence) electrons free from external field, 
its electronic Hamiltonian \cite{RMPxiaoyuan} is expressed in $K$ canonical molecular orbitals as 
\begin{eqnarray}
    \hat{H}_0 &=& \sum_{p,q=1}^{K}h_{pq}a_p^\dagger a_q + \frac{1}{2}
    \sum_{p,q,r,s=1}^{K}h_{p q r s}a_p^\dagger a_q^\dagger a_s a_r, \label{eq:H0}
\end{eqnarray}
where $p, q, r,$ and $s$ denotes different spin orbitals and parameters (one-electron $h_{pq}$ and two-electron $h_{pqrs}$ integrals) 
\begin{eqnarray}
    h_{pq} &=& \int d\mathbf{x}\phi_{p}^{*}(\mathbf{x})\left(-\frac{\nabla^{2}}{2}-\sum_{I=1}^{N}\frac{Z_{I}}{|\mathbf{r}-\mathbf{R_{I}}|}\right)\phi_q(\mathbf{x}) \label{eq:Hpq},\notag\\
    h _{p q r s} &=& \int d\mathbf{x_{1}} d\mathbf{x_{2}}\frac{\phi_p^* (\mathbf{x_1})\phi_q^* (\mathbf{x_2})\phi_s (\mathbf{x_1})\phi_r(\mathbf{x_2})}{|\mathbf{x_{1}} - \mathbf{x_{2}}|}\label{eq:Hpqrs}
\end{eqnarray}
are pre-calculated on classical computers by Hartree-Fock method with $\mathcal O(K^4)$ scaling. 
Using the VQE method to compute ground state energy requires a parameterized ansatz or a trial wave function.
Many different forms of wavefunctions are developed for such a purpose.
A popular ansatz used by VQE method is unitary coupled-cluster ansatz with single and double excitations (UCCSD) \cite{PRXroos},
given by
\begin{widetext}
\begin{equation}
    \begin{aligned}
    |\Psi_{\text{UCCSD}}(\theta)\rangle=e^{\hat{T}_{sd}+\hat{T}^\dagger_{sd}}|\Psi_0\rangle=\exp{\left[\sum_{ra}c_a^r(a_r^\dagger a_a-a_a^\dagger a_r)+\sum_{a<b,r<s}c_{ab}^{rs}(a_r^\dagger a_s^\dagger a_a a_b-a_b^\dagger a_a^\dagger a_s a_r)\right]}|\Psi_0\rangle,
    \end{aligned}
\end{equation}
\end{widetext}
where $\hat{T}_{sd}=\hat{T}_1+\hat{T}_2$ contains single and double excitation operators.
$|\Psi_0\rangle$ is the ground state wave function obtained by Hartree-Fock method.
Minimizing the expectation value of molecular energy $\langle\Psi_{\text{UCCSD}}(\theta)|\hat{H}_{0}|\Psi_{\text{UCCSD}}(\theta)\rangle$,
we find the ground state wave function and the corresponding energy.

For a two-electron system equipped with STO-3G basis,
it has 4 spin orbitals denoted by $g,\bar{g},u,\bar{u}$,
representing gerade spin-up,
gerade spin-down, 
ungerade spin-up,
and ungerade spin-down orbitals,
respectively.
It is evident the two electrons occupy two of the 4 spin orbitals,
and as a result the ground state function of the molecule stay in the linear space $V$ spanned by the following four basis states \cite{arxivfang}
\begin{eqnarray}
    V=\{|\sigma_g\bar{\sigma}_g\rangle, |\sigma_u\bar{\sigma}_u\rangle,
    |\sigma_g\bar{\sigma}_u\rangle, |\sigma_u\bar{\sigma}_g\rangle\}. \label{eq:V}
\end{eqnarray}
For H$_2$ molecule,
the ground state wavefunction $|\psi_{H_2}\rangle$ stays in a smaller linear space
\begin{eqnarray}
    V_g=\{|\sigma_g\bar{\sigma}_g\rangle, |\sigma_u\bar{\sigma}_u\rangle\}\label{eq:Vg}
\end{eqnarray}
due to its central symmetry.
As a result,
the electronic Hamiltonian \eqref{eq:H0} for the two-electron system is simplified into 
\begin{widetext}
\begin{equation}
\begin{aligned}
H=& h_{gg}(a_g^\dagger a_g + a_{\bar{g}}^\dagger a_{\bar{g}}) + h_{\mu\mu}(a_\mu^\dagger a_\mu + a_{\bar{\mu}}^\dagger a_{\bar{\mu}})+h_{gg\mu\mu}(a_g^\dagger a_{\bar{g}}^\dagger a_{\bar{\mu}} a_\mu + a_g^\dagger a_{\bar{\mu}}^\dagger a_{\bar{g}} a_\mu + a_\mu^\dagger a_{\bar{g}}^\dagger a_{\bar{\mu}} a_g + a_\mu^\dagger a_{\bar{\mu}}^\dagger a_{\bar{g}} a_g)\notag\\
&+ (h_{g\mu g\mu}-h_{g \mu \mu g})(a_g^\dagger a_\mu^\dagger a_\mu a_g + a_{\bar{g}}^\dagger a_{\bar{\mu}}^\dagger a_{\bar{\mu}} a_{\bar{g}})+h_{gggg}a_g^\dagger a_{\bar{g}}^\dagger a_g a_{\bar{g}}  + h_{\mu\mu\mu\mu}a_\mu^\dagger a_{\bar{\mu}}^\dagger a_{\bar{\mu}} a_\mu+h_{g\mu g\mu}(a_g^\dagger a_{\bar{\mu}}^\dagger a_{\bar{\mu}}a_g + a_{\bar{g}}^\dagger a_\mu^\dagger a_\mu a_{\bar{g}}).
\end{aligned}
\end{equation}
\end{widetext}

To perform the VQE algorithm,
we need to encode or map ground state wave function and the Hamiltonian into quantum circuits.
There exit several schemes to encode creation and annihilation operators into strings of Pauli gates.
The common choices are Jordan-Wigner,
Parity,
and Bravyi-Kitaev transformation.
In our simulation,
we employ the Bravyi-Kitaev (BK) transformation \cite{JCPseeley} to encode the wavefunctions and Hamiltonians,
which enables the reduction of using two qubits for simulations or experiments through $Z_2$ symmetry.

For the two-electron system,
we store the occupation numbers of spin orbitals $|\Phi_0\rangle$ in the order of $|f_{\bar{\sigma}_u},f_{\bar{\sigma}_g},f_{\sigma_u},f_{\sigma_g}\rangle$,
where $f$ equals to 1 (occupied) or 0 (unoccupied).
In the following,
we denote the order $\sigma_u,\sigma_g,\sigma_u,$ and $\sigma_g$ by $3,2,1,$ and $0$ for simplicity.
As a result,
the indices in Hamiltonian \eqref{eq:H0} and $|\Phi_0\rangle$ are modified.
Under the BK transformation, the encoded wavefunction of two-electron system \eqref{eq:V} stays in the space $V^{\text{qubit}}_g = \{|0101\rangle,|1010\rangle,|0110\rangle,|1001\rangle\}$,
or $= \{|11\rangle,|00\rangle,|10\rangle,|01\rangle\}$ under $Z_2$ reduction.
The encoded wave function \eqref{eq:Vg} of H$_2$ stays in $V^{\text{qubit}}_u = \{|0101\rangle,|1010\rangle\}$,
or $=\{|11\rangle,|00\rangle\}$ under $Z_2$ reduction.

After some calculation,
the UCCSD operator of two-electron system is found to take the following form
\begin{widetext}
\begin{eqnarray}
    U(\theta) = \exp{\left[\theta_{10}(a_1^\dagger a_0-a_0^\dagger a_1)+\theta_{32} (a_3^\dagger a_2-a_2^\dagger a_3)+\theta_{3120} (a_3^\dagger a_1^\dagger a_0 a_2 - a_2^\dagger a_0^\dagger a_1 a_3)\right]}.
\end{eqnarray}
\end{widetext}
Carrying out the BK encoding,
the UCCSD operator becomes
\begin{eqnarray}
    U(\theta) = \exp{\left\{i\left[\frac{\theta_{3120}}{2}(X_1Y_0+Y_1X_0)+\frac{\theta_{10}}{2}Y_0+\frac{\theta_{32}}{2}Y_1\right]\right\}}\notag,
\end{eqnarray}
and the corresponding Hamiltonian becomes 
\begin{equation}
    \begin{aligned}
    H_{\rm{HeH}^+}^{\rm{BK}}=&a_0I+a_1IZ+a_2IX+a_3ZI+a_4XI\\
    	&+a_5ZZ+a_6ZX+a_7XZ+a_8XX.
    \end{aligned}
\end{equation}

The single excitation terms $a_1^\dagger a_0-a_0^\dagger a_1$ and $a_3^\dagger a_2-a_2^\dagger a_3$ are discarded for H$_2$ molecule due to its spatial symmetry,
and the resulting UCCSD operator is simplified into
\begin{eqnarray}
    U(\theta) = \exp{\left[i\frac{\theta}{2}(X_1Y_0+Y_1X_0)\right]}.
\end{eqnarray}
For the ground state of H$_2$ (state $|11\rangle$ under BK transformation with Z$_2$ symmetry),
the ansatz and Hamiltonian are further simplified into
\begin{equation}
    \begin{aligned}
 U(\theta)|11\rangle &= e^{i\theta X_1Y_0}|11\rangle, \\
 H^{\text{BK}}_{\rm{H}_2} &= a_0 I + a_1Z_0 + a_2Z_1 + a_3 Z_1Z_0 + a_4 X_1X_0. \label{eq:uccsd}
    \end{aligned}
\end{equation}
To evaluate the ground state energy of system,
we need to compute the expected value of each term in $H^{BK}_{H_2}$, which is measured directly with quantum circuits.

\subsection{Derivation of UCCSD Ansatz of H$_2$}

\begin{figure}
    \includegraphics[width=3.4in]{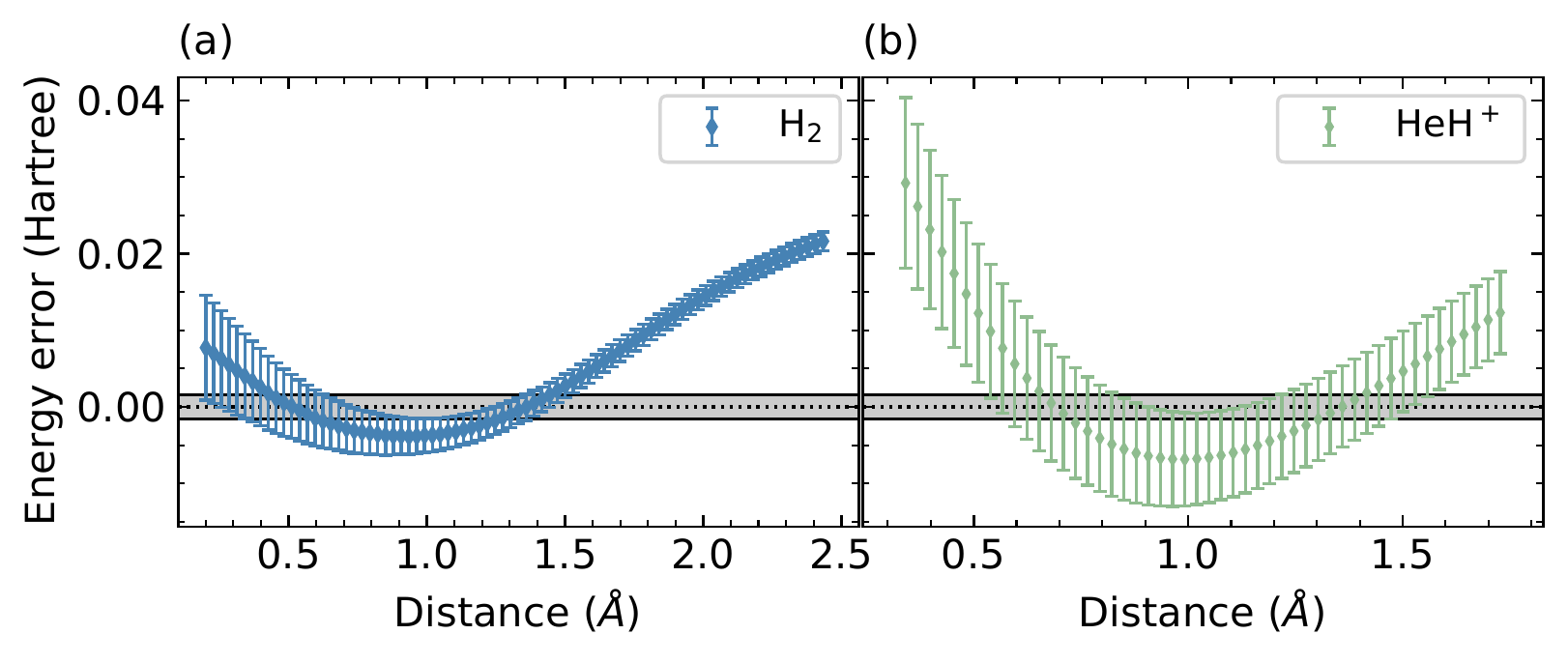}
    \caption{\label{fig:veqerror}
        \textcolor{black}{The energy error and standard deviation for (a) $\rm{H}_2$ and (b) $\rm{HeH}^+$.
        The energy error is defined as the difference between the measured and exact energy,
        with the dotted line the expected error and the gray shadow region denoting the chemical accuracy of $\pm1.6\times10^{-3}$ hartree.
        The error bars represent the standard deviation of the sample mean obtained by bootstrapping from $4096$ projective measurement repetitions.
        The data points for $\rm{H}_2$ are obtained from the Gaussian process regression on 50 different values for the variational parameter.}
    }
\end{figure}

The general form of UCC operator is given by 
\begin{eqnarray}
    U(\theta) = e^{T(\theta)-T^{\dagger}(\theta)}
\end{eqnarray}
where $T(\theta)$ is the excitation operator defined by full configuration interaction (FCI),
and for most of the time,
the UCCSD ansatz is sufficient to deal with this problem.
As mentioned above,
we store the occupation numbers of spin orbitals ordered by $|f_{\bar{\sigma}_u},f_{\bar{\sigma}_g},f_{\sigma_u},f_{\sigma_g\rangle}$.
As a result,
the initial state of H$_2$ (obtained through classical quantum chemistry computation such as Hartree-Fock method) in particle number representation is $|0101\rangle$.
Accordingly,
following the BK transformation,
the initial state changes to $|0111\rangle$ (or $|11\rangle$ under Z$_2$ reduction).
Thus,
for H$_2$ with STO-3G basis,
the operator is simplified into \cite{PRXroos} 
\begin{eqnarray}
U(\theta) = e^{\theta_{3120} (a_3^\dagger a_1^\dagger a_0 a_2 - a_2^\dagger a_0^\dagger a_1 a_3)}.
\end{eqnarray}

Following BK encoding,
the fermionic operators are given by
\begin{equation}
    \begin{aligned}
    a_0^\dagger  &=  \frac{1}{2}X_3X_1X_0 - \frac{1}{2}i X_3X_1Y_0,& \\
    a_0               &=\frac{1}{2}X_3X_1X_0 + \frac{1}{2}i X_3X_1Y_0,&\\
    a_1^\dagger  &=  \frac{1}{2}X_3X_1Z_0 - \frac{1}{2}i X_3Y_1,&\\
    a_1               &=\frac{1}{2}X_3X_1Z_0 + \frac{1}{2}i X_3Y_1,&\\
    a_2^\dagger &=  \frac{1}{2}X_3X_2Z_1 - \frac{1}{2}i X_3Y_2Z_1,&\\
    a_2              &= \frac{1}{2}X_3X_2Z_0 + \frac{1}{2}i X_3Y_2Z_1,&\\
    a_3^\dagger &= \frac{1}{2}X_3Z_2Z_1 - \frac{1}{2}i Y_3,&\\
    a_3              &=\frac{1}{2}X_3Z_2Z_1 + \frac{1}{2}i Y_3.& \label{eq:fermiOp}
    \end{aligned}
\end{equation}
For Pauli matrices (or Pauli gates),
they satisfy the equalities: $XX = YY = ZZ =1, XY = -YX = iZ, YZ = -ZY = iX,$ and $ZX =-XZ = iY$.
Based on the above equations,
it is easy to show that
\begin{equation}
    \begin{aligned}
    &a_3^\dagger a_1^\dagger a_0 a_2 - a_2^\dagger a_0^\dagger a_1 a_3\\
    =\frac{1}{8}i ( &Y_2X_0+X_2Y_0 - Y_2Z_1X_0 + X_2Z_1Y_0+Z_3X_2Z_1Y_0\\
    &- Z_3Y_2Z_1X_0 + Z_3X_2Y_0 + Z_3Y_2X_0 ).
    \end{aligned}
\end{equation}
To omit the qubit 1 and qubit 3 in $|0111\rangle$ by Z$_2$ symmetry,
one needs to consider the effect of pauli matrices acting on qubits 1 and 3,
when the remaining two qubits (i.e., qubit 0 and qubit 4) are concerned.
For the eight terms in the above equation,
only Z matrix acts on the qubits 1 and 3.
At the same time,
it is evident that $Z_3|0111\rangle = |0111\rangle$ and $Z_1|0111\rangle = -|0111\rangle$.
Thus the form of $a_3^\dagger a_1^\dagger a_0 a_2 - a_2^\dagger a_0^\dagger a_1 a_3$ after Z$_2$ reduction becomes
\begin{eqnarray}
    && a_3^\dagger a_1^\dagger a_0 a_2 - a_2^\dagger a_0^\dagger a_1 a_3 = \frac{1}{2}(X_2Y_0+Y_2X_0),
\end{eqnarray} 
which,
equivalently,
can also be expressed as $\frac{1}{2}(X_1Y_0+Y_1X_0)$,
since qubit 1 is omitted (allowing us to renumber qubit 2 as 1).
Therefore,
the ansatz $U(\theta) = e^{\frac{1}{2}i\theta(X_1Y_0+Y_1X_0)}$ is rewritten as $e^{\frac{1}{2}i\theta X_1Y_0}e^{\frac{1}{2}i\theta Y_1X_0}$.
On the other hand, 
$e^{\frac{1}{2}i\theta X_1Y_0}|11\rangle = e^{\frac{1}{2}i\theta Y_1X_0}|11\rangle$ holds,
and as a result the UCCSD ansatz of H$_2$ with STO-3G is recast into
\begin{eqnarray}
U(\theta) = e^{i\theta X_1Y_0}|11\rangle = e^{i\theta Y_1X_0}|11\rangle.
\end{eqnarray}

\textcolor{black}{The results for measured errors for energy with fluctuations for $\rm{H}_2$ and $\rm{HeH}^+$ is respectively shown in Fig.~\ref{fig:veqerror}.}

\end{document}